\documentclass[a4paper, 10pt]{article}
\usepackage[utf8]{inputenc}
\usepackage{cite}
\usepackage{color}
\usepackage{hyperref} 
 \usepackage{latexsym}
  \usepackage{amssymb}
  \usepackage{graphicx}
   \DeclareGraphicsExtensions{.jpg .png .gif .pdf .pdf}
\usepackage[utf8]{inputenc}
\usepackage[T1]{fontenc}
 \usepackage[english]{babel}
  \usepackage{}
  \usepackage{amstext}
  \usepackage{amsmath}
  \usepackage{times} 
  \usepackage{amsfonts}
 \date{}

 \title{ Linear Stabily analysis of hydromagnetic Couette flow  with small 
 injection/suction through the modified Orr-Sommerfeld  equation}
 \author{L. A. Hinvi\footnote{laurent.hinvi@imsp-uac.org}\hspace{0.08cm} 
 A. V. Monwanou\footnote{movins$2008$@yahoo.fr} \hspace{0.05cm}  and J. B. Chabi
   Orou\footnote{Author to whom correspondence should be addressed :  jchabi@yahoo.fr}}
 \begin{document}
 \maketitle{Institut de Math\'ematiques et de Sciences Physiques,  BP :  613 Porto Novo,  B\'enin}
 \begin{abstract}
 This paper analyses  the effects of  small injection/suction Reynolds' number,  
 Hartmann number,  permeability parameter and wave number on a viscous  incompressilbe electrically 
 conduction   fluid flow in a parallel porous channel. The plates of the  channel with  small constant 
 injection/suction,   have constant temperature. The upper plate 
 is allowed to mouve  in flow direction and the lower plate is kept at rest.
 A magnetic field   of uniform strength 
 is also applied normally to the plates what are parallel.  The originality of the paper is 
 to study the effect of the above parameter in temporal linear stabilty analysis of the flow through 
 the modified Orr-Sommerfeld equation.
 \end{abstract}
 \section{Introduction}

 The study of Couette flow in a rectangular channel of an
electrically conducting viscous fluid under the action of a
transversely applied magnetic field has immediate applications in many devices such as magnetohydrodynamic
(MHD) power generators,  MHD pumps,  accelerators, 
aerodynamics heating,  electrostatic precipitation,  polymer technology,  petroleum industry,  purification of
crude oil and fluid droplets sprays. Channel flows of a
Newtonian fluid with heat transfer were studied with or
without Hall currents by many authors \cite{1,  2,  3, 4, 5, 6, 7, 8, 9, 10, 11}.
The effects of  :  injection/suction through the injection/suction parameter number, 
Hartmann number,  Permeability or Darcy parameter \dots on the stability of the fluids flows were studied
by many researchers \cite{3, 4, 5, 6, 7} with different approch.

 The heat source and the Soret effects on  an oscillatory hydromagnetic flow through a 
porous medium bounded by two vertical parallel porous plates is
analyzed by K. Chand  et al.\cite {3},  where one plate of the 
channel is kept stationary and the other is moving with uniform velocity. 
The plates of the channel are subjected to constant injection and suction velocities respectively.
A. NAYAK et al.\cite{4} have studied  an oscillatory effects on magneto-hydrodynamic flow and heat transfer in rotating
horizontal porous channel.
N. V. R. V. Prasad et al. \cite{5}  have  considered the unsteady hydromagnetic incompressible viscous fluid flow
through a porous medium in a horizontal channel under prescribed discharge,  under the influence of
inclined magnetic field.
 S.S. DAS  et al. \cite{6} analyze the effects of constant suction and sinusoidal injection on
three dimensional couette flow of a viscous incompressible electrically conducting fluid through a
porous medium between two infinite horizontal parallel porous flat plates in presence of a transverse
magnetic field. The stationary plate and the plate in uniform motion are,  respectively,  subjected to a
transverse sinusoidal injection and uniform suction of the fluid.
Shalini et al.\cite{7} have studied the  effects of
variable viscosity and heat source on unsteady laminar flow of dusty conducting 
fluid between parallel porous plates through porous medium with temperature
dependent viscosity.It is assumed that the parallel plates are porous and subjected
to a uniform suction from above and injection from below.

The objective of the present paper is to study the effects of 
the injection/suction Reynods' number, 
Hartmann number,  Permeability  parameter on the linear temporal stability  of the flow  of a viscous
incompressible electrically conducting fluid in a Couette
horizontal porous channel in the presence of a uniform transverse magnetic field when it is fixed
relative to the fluid through the  modified Orr-Sommerfeld equation .
The plates of the channel are considered porous and
flow within the channel is due to the  uniform motion of the upper plate.
Such  linear temporal stability  analysis through 
a modified Orr-Sommerfeld equation has been made earlier by  
A. V. Monwanou and J. B. Chabi Orou \cite{2} in 
a Poiseuille flow without injection/suction. The same problem with 
a viscous incompressible no-conduction fluid
but with small injection/ suction in medium through the porous
plates fat has been made by L.hinvi et al. \cite{9}.The paper is organized as follows.

In the second section the modified Orr-Sommerfeld equation governing the 
stability analysis in the    Couette horizontal 
porous  plates  flow is checked. In the third
section  an  analysis of  the effects of  small injection/suction Reynolds' number $R_{e\omega}$,  
 Hartmann number$M$ and permeability parameter $K_{p}$ on a viscous  incompressilbe electrically 
 conduction   fluid flow in a parallel porous 
 channel linear  stability will be invertigated with the help of figures and tables.
. The conclusions is presented in the final section.



 \section{Modified Orr-Sommerfeld equation}

 We considered a Poiseuille  viscous incompressilbe and
 electrically conduction fluid flow between 
 two porous parallel plates of infinite lengh,  distant $h$ apart in the presence of uniform 
 transverse constant magnetic field $B_{0}$ applied parallel to $y^{*}$ axis which is normal to 
 the planes  of the plates. We considered the simple case where,  $B_{0}$ is  fixed  relative to the fluid.
 We work at constant temperature,  the heat transfert aspect of the 
 flow is not studied. We applied a small constant 
 injection $V_{\omega}$,  at the lower plate and a same small  constant suction $V_{\omega}$,  
 at the upper plate.The upper plate is allowed to mouve with non-zero uniform velocity $U=U_{0}$ in flow
 direction and the lower plate is kept at rest. We choose the origine on the plane $(x^{*}, 0,  z^{*})$ 
 such as $-{h}\leq y^{*}\leq {h}$ and $x^{*}$ parallel to the direction of 
 the motion of the upper plate. We assumed the magnetic Reynolds' number very small
 for metallic liquids and neglected   the induced magnetic field
 in comparison with the applied one \cite{4, 7, 8}. Initially,  $t^{*}<0$,  both the fluid
 and plates are assumed to be at rest. 
 When $t^{*}>0$,  the upper plate starts moving  with a constant velocity $U$  in
 coordinate system with the fluid. The equations of continuity,  motion
 for the viscous incompressible electically conduction fluid in vector form are : 
 \begin{eqnarray}
 \triangledown .V\label{eq1}\\
 \frac{\partial{V}}{\partial{t^{*}}} +(V.\triangledown)V &=& -\frac{1}{\rho}\triangledown{p}+
 \nu {\triangledown}^{2}V +\frac{1}{\rho} J\wedge B-\frac{\mu V}{\rho k^{*}},  \label{eq2}\\
  \triangledown \wedge B&=&{\mu}_{e}J \label{eq3}, \\
  \triangledown \wedge E &=& \frac{\partial{B}}{\partial{t^{*}}},  \label{eq4}\\
  {\triangledown.}B=0,  \label{eq5}\\
  {\triangledown.}J=0.\label{eq6}
 \end{eqnarray}
 They are   continuity,  Newton's second Law,  Ampere's Law,  Faraday's Law, 
 Maxwell's Law and  Gauss Law 
 equations respectively,  with 
 \begin{eqnarray}
 J=\sigma \left(E+V\wedge B\right). \label{eq7}
 \end{eqnarray}
 Where $V\left(u^{*},  v^{*},  w^{*}\right)$,  $B$,  $E$,  $J$,  $\sigma$,  
 ${\mu}_{e}$ the velocity,  the magnetic field,  the electric field 
 , the current density vector,  the fluid electrical conductivity and the magnetic 
 permeability of the fluid respectively and  $t^{*}$ denotes time.
 The physical model of the problem is illustrated in figure \ref{fig : 0}  below,  
 where $V\left(u^{*},  v^{*},  w^{*}\right)$
 is the velocity vector in the $x^{*},  y^{*},  z^{*}$ directions respectively.
 
\begin{figure}[htbp]
 \begin{center}
 \includegraphics[width=6cm]{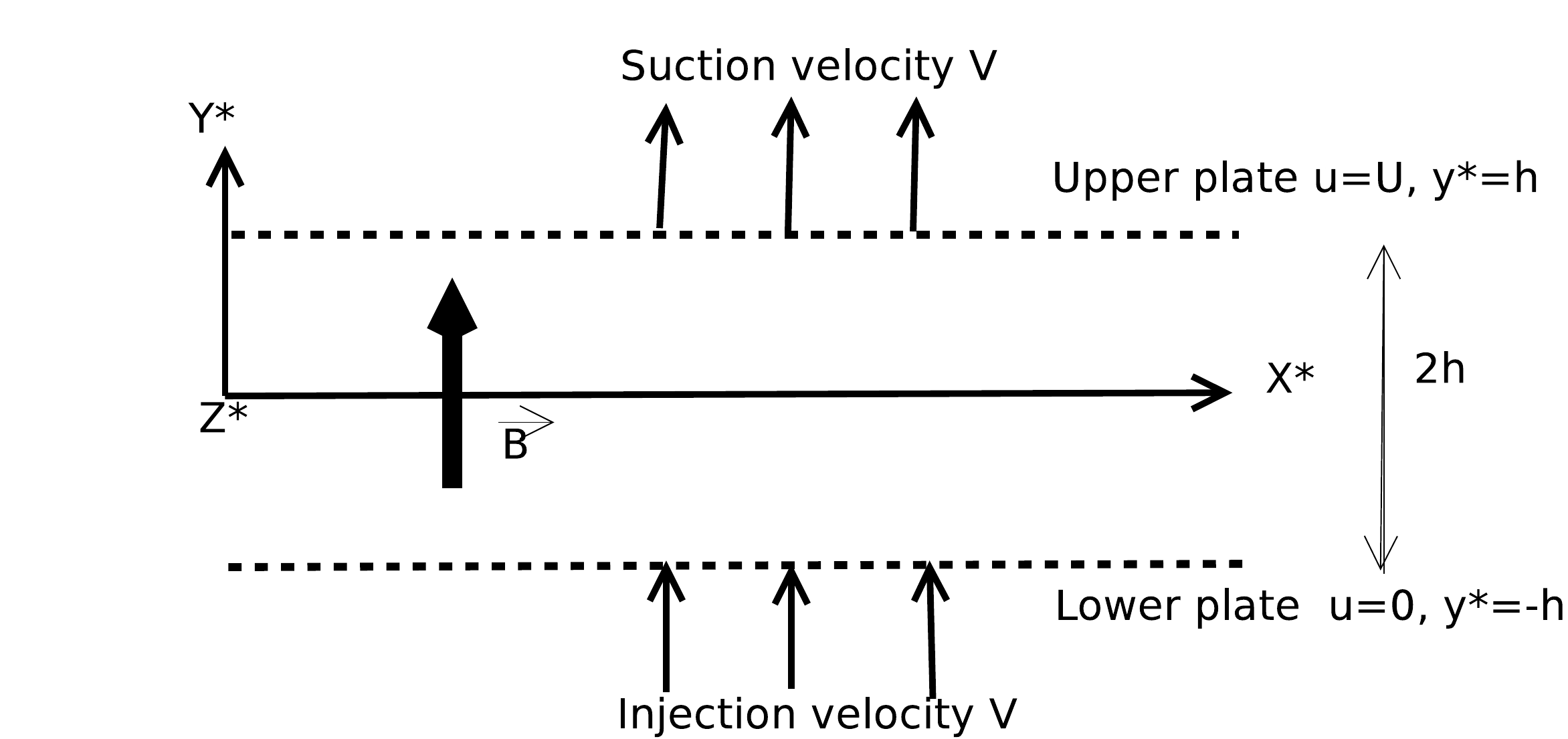} 
 \end{center}
 \caption{physical model and coordinate system } 
 \label{fig : 0}
 \end{figure}
 \begin{eqnarray}
 B&=&\left(0,  B_{0},  0\right), \label{eq8}\\
 E&=&\left(E_{x},  E_{y}, E_{z}\right), \label{eq9}\\
 J&=&\left(J_{x},  0,  J_{z}\right) ; \label{eq10}
 \end{eqnarray}
 where $B_{0}$ is a constant. We assumed that no applied and 
 polarization voltage exists (i.e.,  $E=0$). Then
 \begin{eqnarray}
  (\ref{eq7}) and (\ref{eq10}) \Longrightarrow J=\sigma B_{0}\left(-w,  0,   u\right)\label{eq11}
 \end{eqnarray}
 and (\ref{eq6})  yields
 \begin{eqnarray}
 \sigma B_{0} \left(\frac{\partial{u}}{\partial{z}}
 -\frac{\partial{w}}{\partial{x}} \right)
 &=&0\label{eq12}
\end{eqnarray}
 We introduced the following non-dimensional quantities  
 $\tilde{x}=\frac{x^{*}}{h}$,  $\tilde{y}=\frac{y^{*}}{h}$,  $\tilde{z}=\frac{z^{*}}{h}$, 
 $\tilde{t}=\frac{Ut^{*}}{h}$,   $\tilde{u}=\frac{u^{*}}{U}$, 
 $\tilde{v}=\frac{v^{*}}{V_{\omega}}$,  
 $\tilde{w} =\frac{w^{*}}{U}$,  $\tilde{p}=\frac{p^{*}}{\rho U^{2}}$,  
 $R_{e}=\frac{U h}{\nu}$ ( hydrodynamic  Reynolds' number)
 $R_{e\omega}=\frac{V_{\omega} h}{\nu}$ (Reynods' injection/suction number),  
 $M =B_{0}h \sqrt{\frac{\sigma}{\mu}}$
 (Hartmann number),  $K_{p}=\frac{k^{*}}{h^{2}}$ (permeability parameter) 
 and the equations (\ref{eq1}) and (\ref{eq2}) become

 \begin{eqnarray}
 \frac{\partial{\tilde{u}}}{\partial{\tilde{x}}}+
 \frac{R_{e\omega}}{R_{e}}\frac{\partial{\tilde{v}}}{\partial{\tilde{\tilde{y}}}}+
 \frac{\partial{\tilde{w}}}{\partial{\tilde{z}}}&=&0 \label{eq13}
 \end{eqnarray}
 \begin{eqnarray}
  \frac{\partial{\tilde{u}}}{\partial{\tilde{t}}}+
  \tilde{u} \frac{\partial{\tilde{u}}}{\partial{\tilde{x}}}+
  \frac{R_{e\omega}}{R_{e}} \tilde{v}\frac{\partial{\tilde{u}}}{\partial{\tilde{y}}}+
  \tilde{w}\frac{\partial{\tilde{u}}}{\partial{\tilde{z}}} &=&
  -\frac{\partial{\tilde{p}}}{\partial{\tilde{x}}}+
  \frac{{\triangledown}^{2}\tilde{u}}{R_{e}}-
  \frac{M^{2}}{R_{e}}\tilde{u}-\frac{\tilde{u}}{K_{p}R_{e}} \label{eq14}, \\
    \frac{\partial{\tilde{v}}}{\partial{\tilde{t}}}+
    \tilde{u} \frac{\partial{\tilde{v}}}{\partial{\tilde{x}}}
 +\frac{R_{e\omega}}{R_{e}} \tilde{v}\frac{\partial{\tilde{v}}}{\partial{\tilde{y}}}+
  \tilde{w}\frac{\partial{\tilde{v}}}{\partial{\tilde{z}}} &=&
  -\frac{R_{e}}{R_{e\omega}}\frac{\partial{\tilde{p}}}{\partial{\tilde{y}}}+
  \frac{{\triangledown}^{2}\tilde{v}}{R_{e}}-\frac{\tilde{v}}{K_{p}R_{e}} \label{eq15}, \\
   \frac{\partial{\tilde{w}}}{\partial{\tilde{t}}}+
   \tilde{u} \frac{\partial{\tilde{w}}}{\partial{\tilde{x}}}+
  \frac{R_{e\omega}}{R_{e}} \tilde{v}\frac{\partial{\tilde{w}}}{\partial{\tilde{y}}}+
  \tilde{w}\frac{\partial{\tilde{w}}}{\partial{\tilde{z}}} &=&
  -\frac{\partial{\tilde{p}}}{\partial{\tilde{z}}}+
  \frac{{\triangledown}^{2}\tilde{w}}{R_{e}}-\frac{M^{2}}{R_{e}}\tilde{w}-
  \frac{\tilde{w}}{K_{p}R_{e}} \label{eq16}.
 \end{eqnarray}
 For the stability analysis,  the flow is decomposed into the mean flow and 
 the disturbance according to
 \begin{eqnarray}
 \tilde{u}_{i}(r, t)&=& U_{i}(r)+u_{i}(r, t), \label{eq17}\\
 \tilde{p}(r, t)&=&P(r)+ p(r, t). \label{eq18}
\end{eqnarray}
studied and analyzed with the help of figures and tables.
 We take  the   dimensional  base flow  for small suction and injection  \cite{1, 10}
 \begin{eqnarray}
  U^{*}(y) &=&\frac{U}{2}\left(\frac{y^{*}}{h} +1\right)\label{eq19} \\
  V^{*}&=& V_{\omega}\label{eq20}\\
  W^{*}&=&0.\label{eq21}
 \end{eqnarray}
 By scaling  these velocities as above,  we obtain with $h^{}=\pm 1$ ($-1\leq y^{*}\leq 1$) 
 the no-dimensional base flow  
 \begin{eqnarray}
  U(y) &=&\frac {y+1}{2} \label{eq22}\\
  V&=& 1\label{eq23}\\
  W&=&0.\label{eq24}
 \end{eqnarray}
 To obtain the stability equations for the spatial evolution of three-dimensional,  
 we take the dependent on time  disturbances 
\begin{eqnarray}
 (u(x,  y,  z,  t) ; v(x,  y,  z,  t) ;   w(x,  y,  z,  t) ;  p(x,  y,  z,  t)) ; \label{eq25}
\end{eqnarray}
 which are  scaled in the same way as above.\\
 We replace the equations  (\ref{eq17}) - (\ref{eq25})  in the equations  (\ref{eq14})-(\ref{eq16}), 
 after linearization and neglection of quadratic terms we find
 
  \begin{eqnarray}
  \frac{\partial{u}}{\partial{t}}+U \frac{\partial{u}}{\partial{x}}+
  \frac{R_{e\omega}}{R_{e}} \frac{\partial{u}}{\partial{y}}+
 \frac{R_{e\omega}}{R_{e}}u \frac{\partial{U}}{\partial{y}}
  &=&-\frac{\partial{p}}{\partial{x}}+
  \frac{{\triangledown}^{2}u}{R_{e}}-\frac{M^{2}}{R_{e}}u-\frac{u}{K_{p}R_{e}} \label{eq26}, \\
    \frac{\partial{v}}{\partial{t}}+U \frac{\partial{v}}{\partial{x}}+
  \frac{R_{e\omega}}{R_{e}} \frac{\partial{v}}{\partial{y}}
   &=&-\frac{R_{e}}{R_{e\omega}}\frac{\partial{p}}{\partial{y}}+
  \frac{{\triangledown}^{2}v}{R_{e}}-\frac{v}{K_{p}R_{e}} \label{eq27}, \\
   \frac{\partial{w}}{\partial{t}}+U \frac{\partial{w}}{\partial{x}}+
  \frac{R_{e\omega}}{R_{e}} \frac{\partial{w}}{\partial{y}}
   &=&-\frac{\partial{p}}{\partial{z}}+
  \frac{{\triangledown}^{2}w}{R_{e}}-\frac{M^{2}}{R_{e}}w-\frac{w}{K_{p}R_{e}} \label{eq28}.
 \end{eqnarray}

  The pressure terms can be eliminated from Navier-Stokes equations.
  For such a mean profile (base flow),  the divergence  of Navier-Stokes  equations
  and continuity,  gives
  \begin{eqnarray}
   {\bigtriangledown}^{2}{p}&=& -2\frac{R_{e\omega}}{R_{e}}\frac{dU}{dy} 
 \frac{\partial{v}}{\partial{x}}+M^{2}\frac{R_{e\omega}}{{R_{e}}^{2}}\frac{\partial{v}}{\partial{y}}.
 \label{eq29}
 \end{eqnarray}
 
 The equations ${\bigtriangledown}^{2}(\ref{eq27})$ and (\ref{eq29})  after linearization give
 \begin{eqnarray}
  \left[ \frac{\partial{}}{\partial{t}}  + 
  {U} \frac{\partial{}}{\partial{x}}+
   \frac{R_{e \omega}}{R_{e}}    
   \frac{\partial{}}{\partial{y}} +\frac{{1}}{K_{p}R_{e}} 
   -\frac{{\bigtriangledown}^{2}}{R_{e}} \right] {\bigtriangledown}^{2}{v} -
   \frac{d^{2}U}{{dy}^{2}}\frac{\partial{v}}{\partial{x}}+
  \frac{M^{2}}{R_{e}}\frac{\partial^{2}{v}}{{\partial{y}}^{2}} 
  &=&0.
 \label{eq30}
\end{eqnarray}
 The disturbances are taken 
 to be periodic in the streamwise,  spanwise directions and time,   
 which allow us to assume solutions of the form
 \begin{eqnarray}
 f(x, y, z, t)&=&\hat{f}(y)e^{i(\alpha x+\beta z-\omega t)} ;  \label{eq31}
 \end{eqnarray}
 where $f$ represents either one of the disturbances $u$,  $v$,  $w$ or $p$ and $\hat{f}$ 
 the amplitude function,  $k$,  $\alpha= k_{x}=k\cos{\theta}$ and  
 $\beta= k_{y}=k\sin{\theta}$ are the wave numbers,  $\omega =\alpha c$ 
 the pulsation of the wave.
 With  $i^{2}=-1$,   $\theta= (\vec{k_{x}},  \vec{k})$,  $c=c_{r}+ic_{i}$ wave velocity  
 which is taken to be complex,  $\alpha$ and $\beta$ are
 real because of temporal stability analysis consideration.
 Then with  the equation (\ref{eq31}),  the equation (\ref{eq30}) becomes
 \begin{eqnarray}
  i\alpha \left(U-c - i\frac{R_{e \omega}D}{{\alpha}R_{e}}- 
  \frac{i}{{\alpha}K_{p}R_{e}}+i\frac{D^{2} -{k}^{2} }{\alpha R_{e}}
  \right) \left(D^{2}-{k}^2  \right)\hat{v}&=& \nonumber\\ 
  - \left(\frac{M^{2}D^{2}}{R_{e}}-i\alpha U''\right) \hat{v}   
  ; \label{eq32}
 \end{eqnarray}
 where $D=\frac{d}{dy}$  ;  with boundary conditions for all $(x, \pm1,  z , t>0)$
 \begin{eqnarray}
 \hat{u}(\pm 1)&=&1 or 0 \label{eq33}\\
    \hat{v}(\pm 1)&=&1 \label{eq34}\\
    \hat{w}(\pm 1)&=&0\label{eq35}\\
    \hat{u}'(\pm 1)&=&0 \label{eq36}\\
    \hat{v}'(\pm 1)&=&0. \label{eq37}\\
     \hat{w}'(\pm 1)&=&0\label{eq38}
  \end{eqnarray}
 Taking
  \begin{eqnarray}
 v_{p}(x, y, z, t)&=& \hat{v}(y)e^{i(\alpha x+\beta z-\omega t)}-1, \label{eq39}
 \end{eqnarray}
  the equation (\ref{eq32}),  the boundary conditions   (\ref{eq34}) and (\ref{eq37}),  take the forms

 \begin{eqnarray}
  \left(U - i\frac{{\bf{R_{e \omega}}}D}{{\alpha} R_{e}}-  \frac{i}{{\alpha}{\bf{K_{p}}}R_{e}}
 +i \frac{D^{2} -{k}^{2} }{\alpha R_{e}} \right) \left(D^{2}-{k}^2  \right)\hat{v}_{p}  \nonumber\\
 - \left( U''+\frac{i{\bf{M^{2}}}D^{2}}{\alpha R_{e}}\right)\hat{v}_{p}   
 &=&   \nonumber\\
 c\left(D^{2}-{k}^2 \right)\hat{v}_{p} ; \label{eq40}
 \end{eqnarray}

 \begin{eqnarray}
     \hat{v}_{p}(\pm 1)&=&0 \label{eq41}\\
     \hat{v}_{p}'(\pm 1)&=&0. \label{eq42}
   \end{eqnarray}
  The equation (\ref{eq40})  
  is a flow equation modified by the  small injection/suction Reynolds number 
  ${R_{e \omega}}$,  the Hartmann number $(M =B_{0}h \sqrt{\frac{\sigma}{\mu}})$, 
  and permeability parameter ($K_{p}=\frac{k^{*}}{h^{2}}$)
  which we call modified Orr-Sommerfeld equation,   rewritten as an eigenvalue problem,  where
  $c$ is the eigenvalue and $\hat{v}_{p}$ the eigenfunction.
  \begin{eqnarray}
   \left[  \left(U - i\frac{{\bf{R_{e \omega}}}D}{\alpha R_{e}}-  \frac{i}{{\alpha}{\bf{ K_{p}}}R_{e}}
 +i \frac{D^{2} -{k}^{2} }{\alpha R_{e}} \right) \left(D^{2}-{k}^2  \right) 
 - U''-\frac{i{\bf{M^{2}}}D^{2}}{\alpha R_{e}} \right]\nonumber
  \end{eqnarray}

   and $\left( D^{2}-k^{2}\right)$ are the operators.

 \section{ Linear Stability analysis}
  We  consider our three-dimensional disturbances. We use a temporal 
 stability analysis as mentioned above. With $c$ complex as we have defined above, 
 when $c_{i}< 0$ we have stability,  $c_{i}=0$ 
 we have neutral stability and  elsewhere we have  instability.
  We employ Matlab $7.8.0.(R2009a)$  in all our numerical computations to find the eigenvalues.
 The Couette  horizontal  porous  plates  flow with the basic velocity profile  
 \begin{eqnarray}
 {\bf{U}} = \left(\frac {y+1}{2},  1,  0 \right) \label{eq43}
 \end{eqnarray}
 for $R_{e\omega}$ small ( i.e. small suction)  is considered.
The eigenvalue problem (\ref{eq40}) is  solved numerically with the suitable boundary conditions.
The solutions are found in a layer bounded at $y=\pm1$ with ${\bf{U}}(\pm1)= (0,  1,  0)$.
The results of calculations are presented in the figures below.
We present the figures related to the eigenvalue problem (\ref{eq40}).

For all these  figures the black,  red,  green  and blue colors are respectively,  
the curves  $ I,  II,  III,  IV$ and the   yellow color is for $c_{i}=0$.
For all frame  $a,  b,  c,  d$ we have fixed free parameters and  we gave  
the curves  $C_{i} $vs.$ R_{e}$ for sequential 
values of  other parameters.

\begin{figure}[htbp]
 \begin{center}
 \includegraphics[width=6cm]{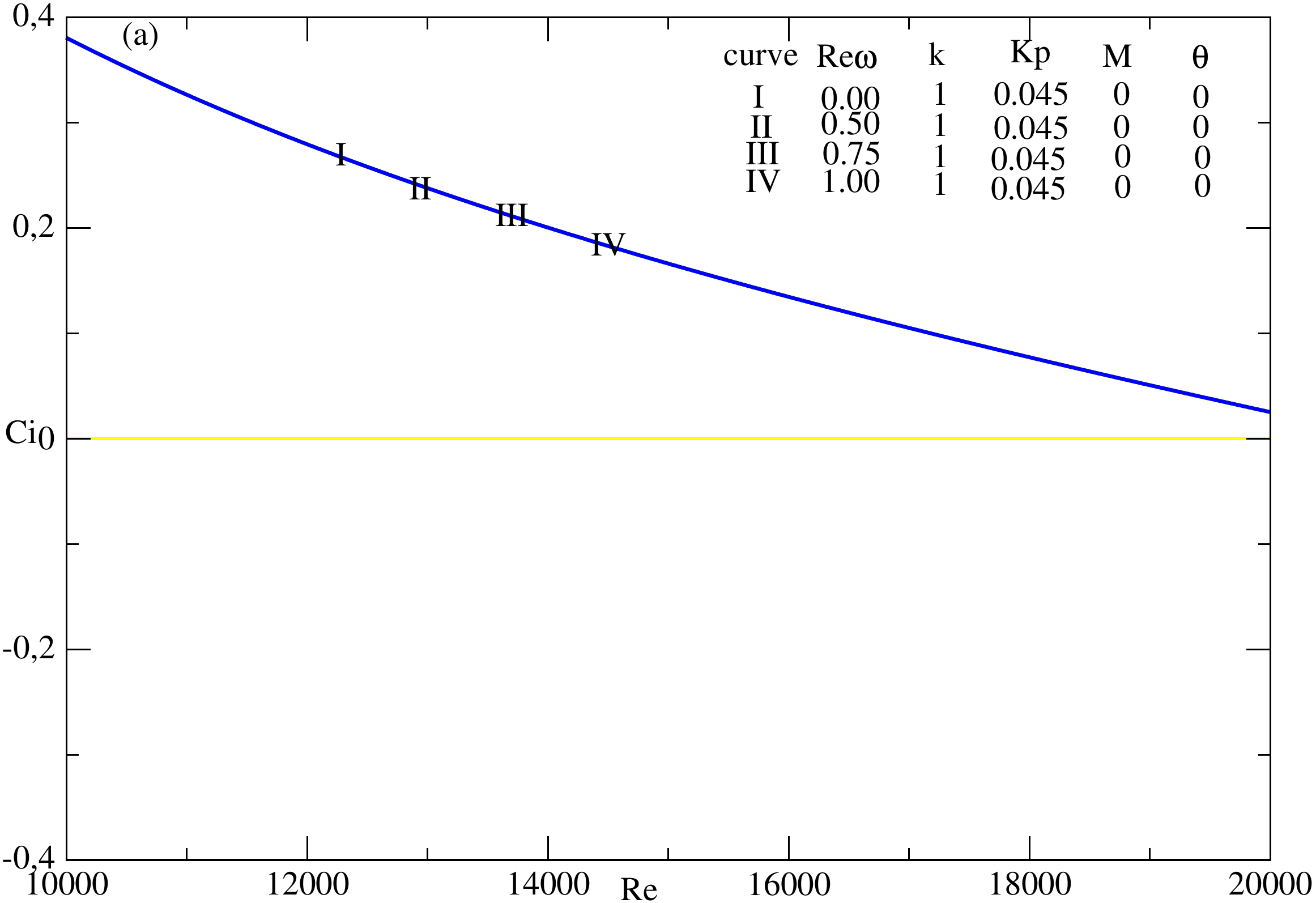} \includegraphics[width=6cm]{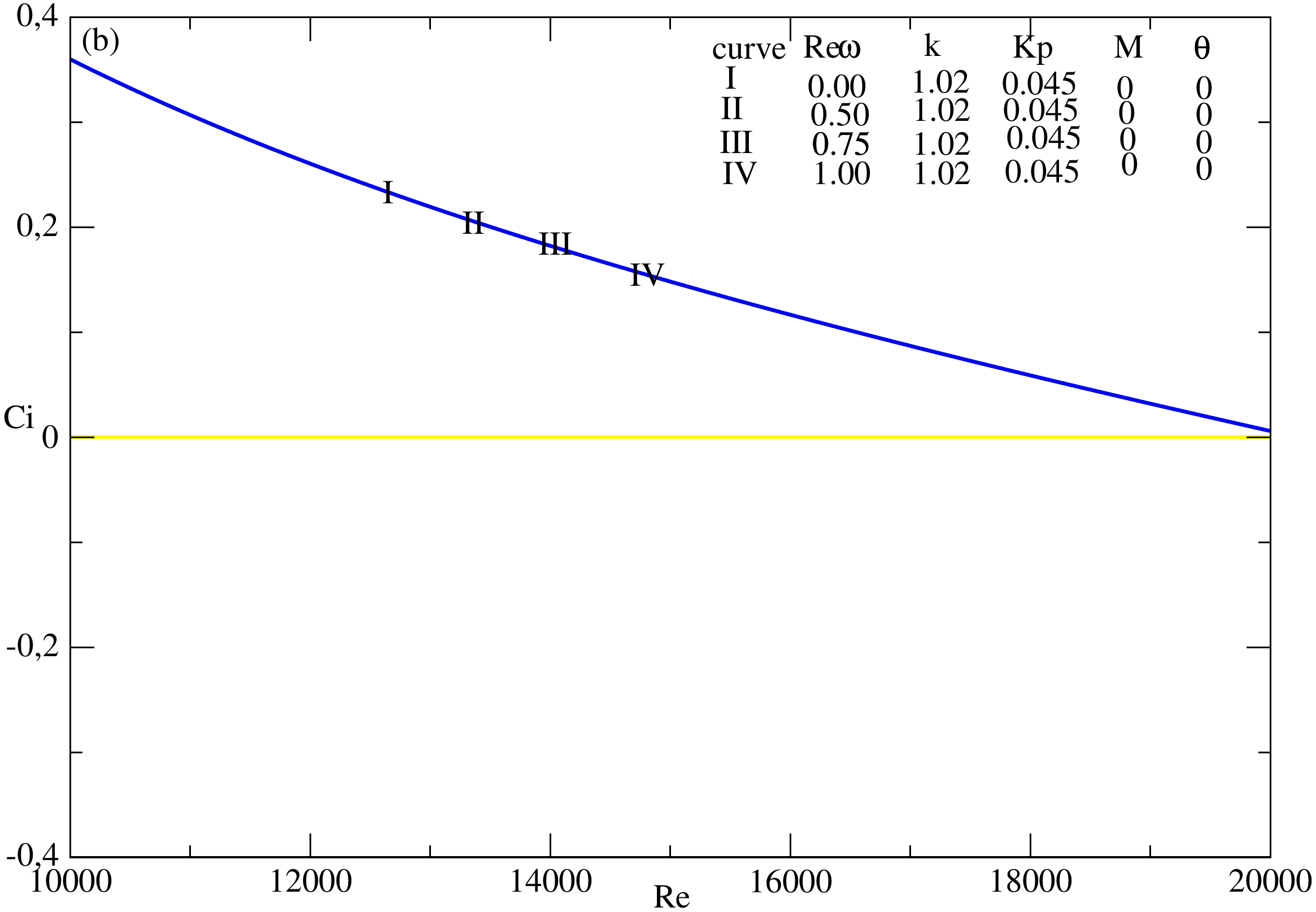} \\
 \includegraphics[width=6cm]{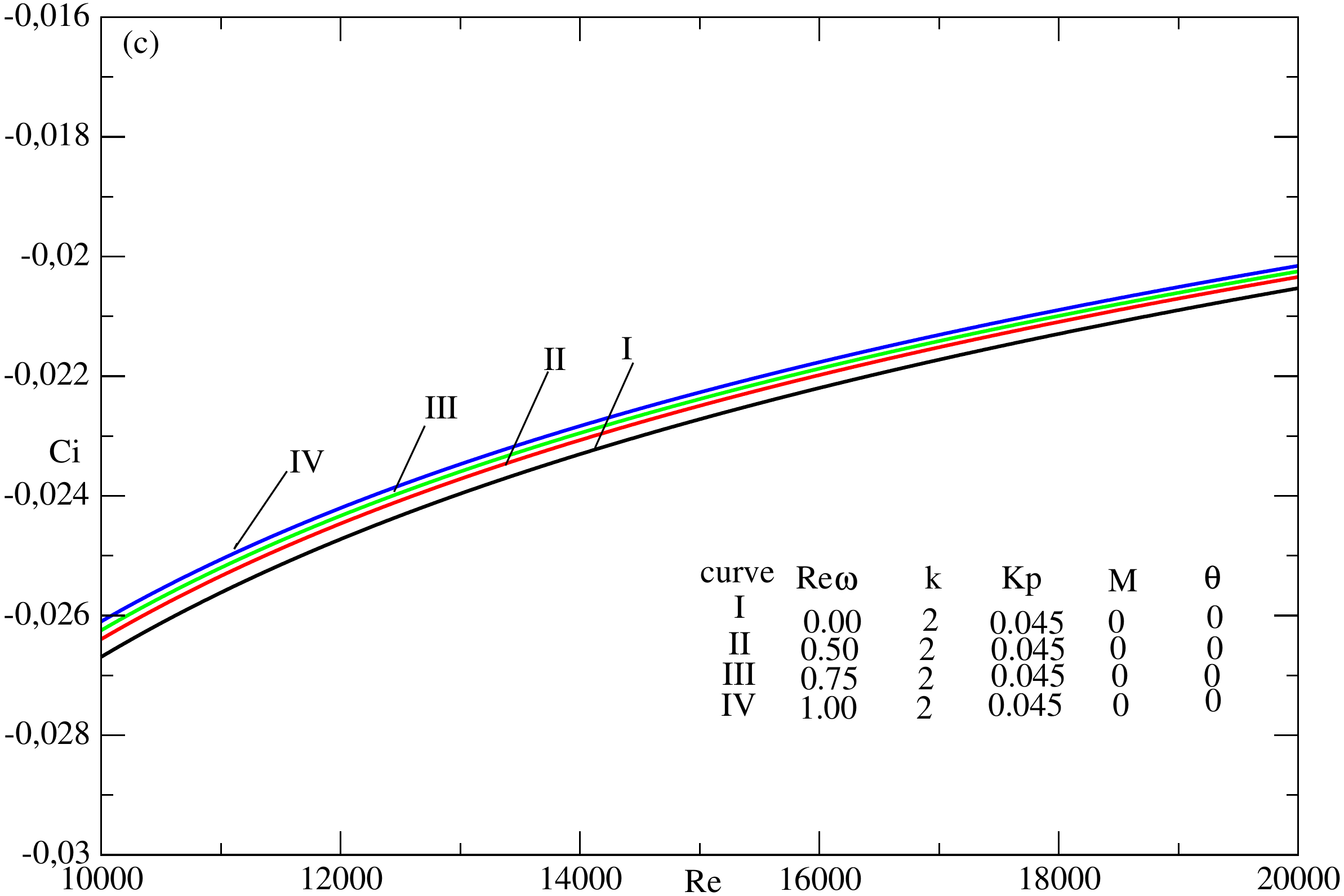} \includegraphics[width=6cm]{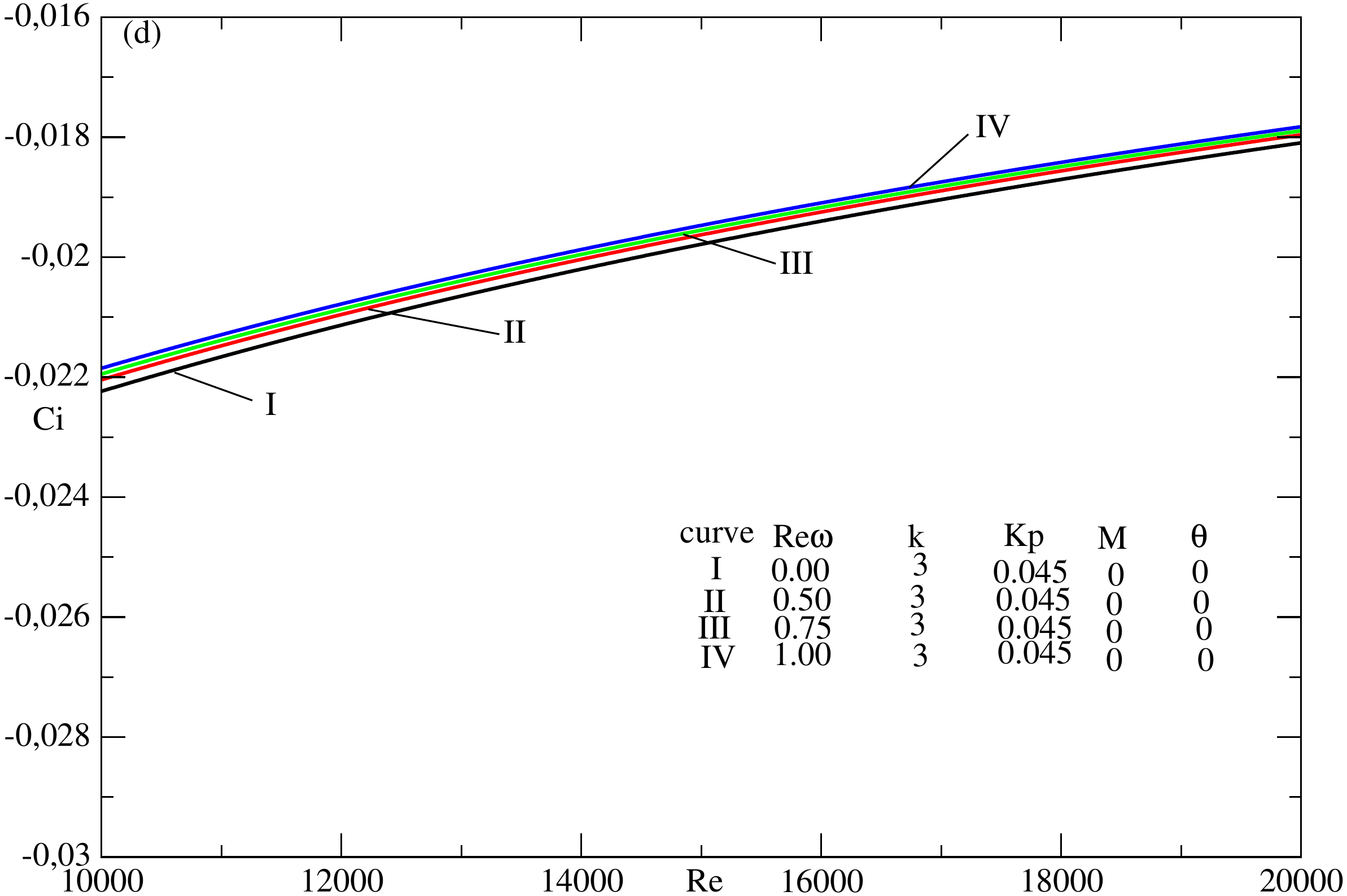}
 \end{center}
 \caption{Ci vs.Re for $M,  K_{p},   \theta$ fixed and  $R_{e\omega},  k$ variable  } 
 \label{fig : 1}
 \end{figure}

 Figure  \ref{fig : 1} presents the effect of  Reynods' injection/suction number $R_{e\omega}$  
 on linear temporal stability of viscous incompressible   
  non electrically conduction fluid ($M=0$) flow for  different values
 of  wave number. It is observed that  for $k=1$ and $k=1.02$ 
 (see figure \ref{fig : 1}  frames a and b),  the stability is not affected by 
 $R_{e\omega}$  and the flow is instable but for $k=2$ and $k=3$ 
 (see figure \ref{fig : 1} frames  c and d), 
 $R_{e\omega}$ affecte it and the flow stays stable,  
 also increase of $R_{e\omega}$ doesn't contribute to  the satability. 
 
\begin{figure}[htbp]
 \begin{center}
 \includegraphics[width=6cm]{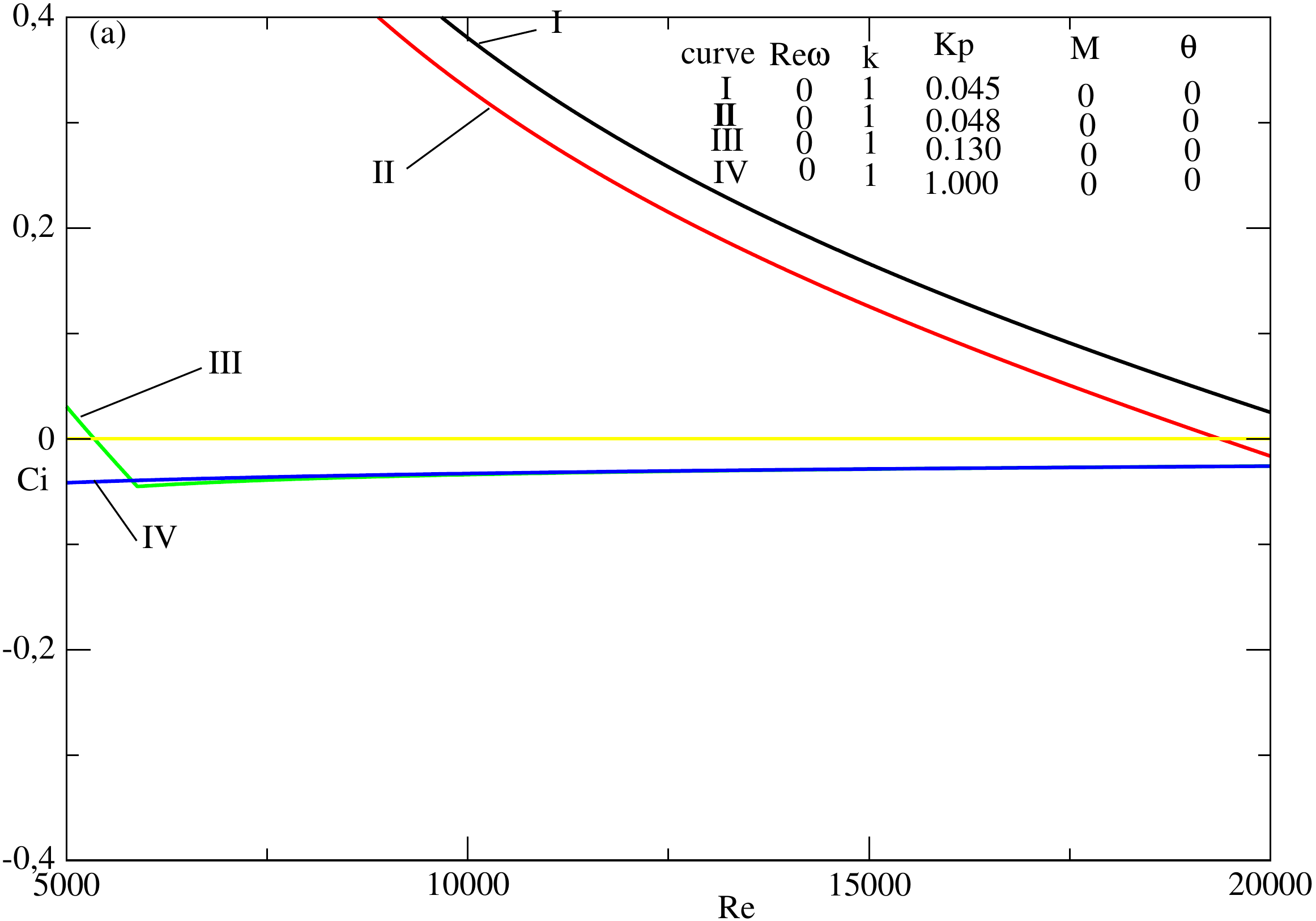} \includegraphics[width=6cm]{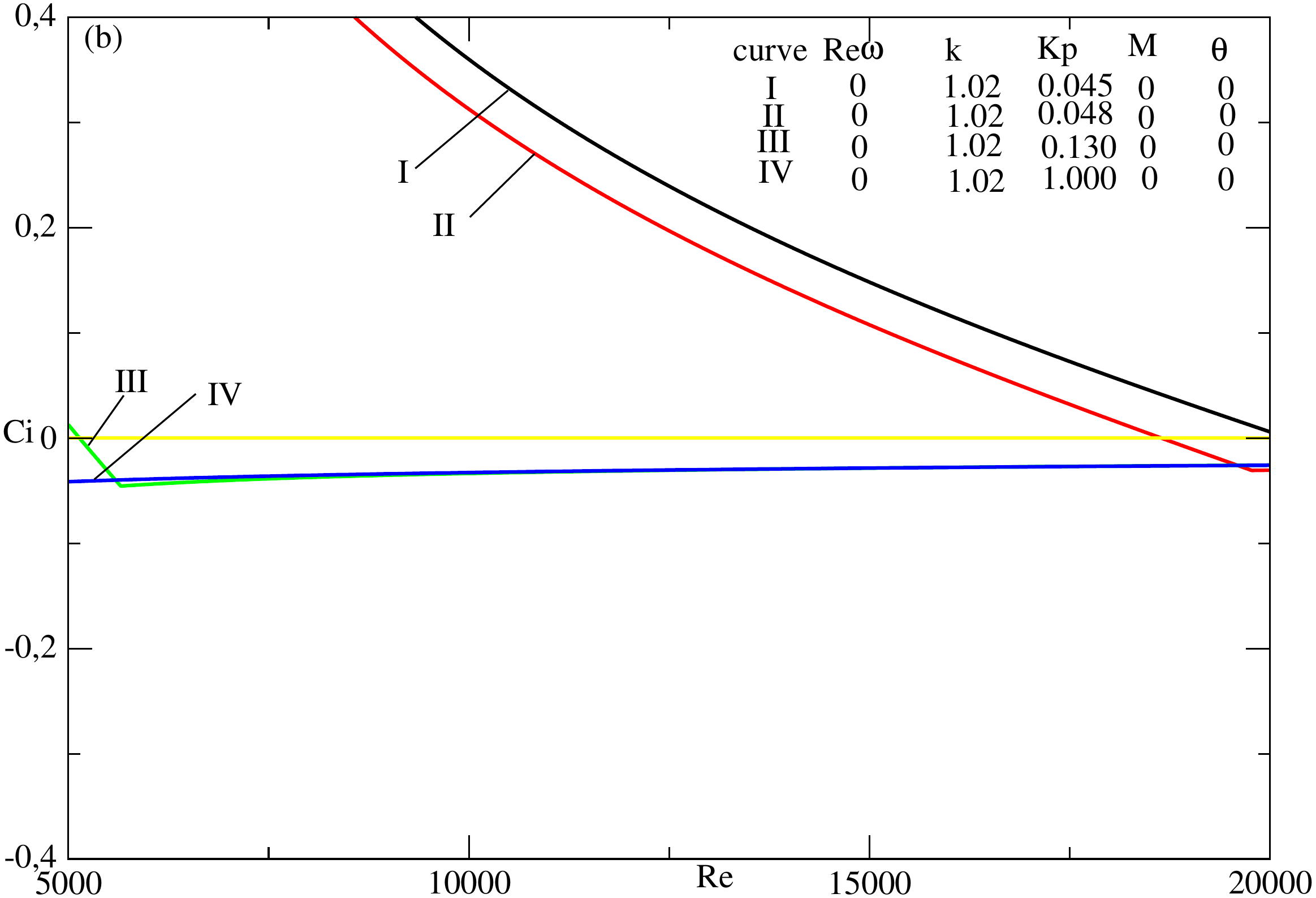} \\
 \includegraphics[width=6cm]{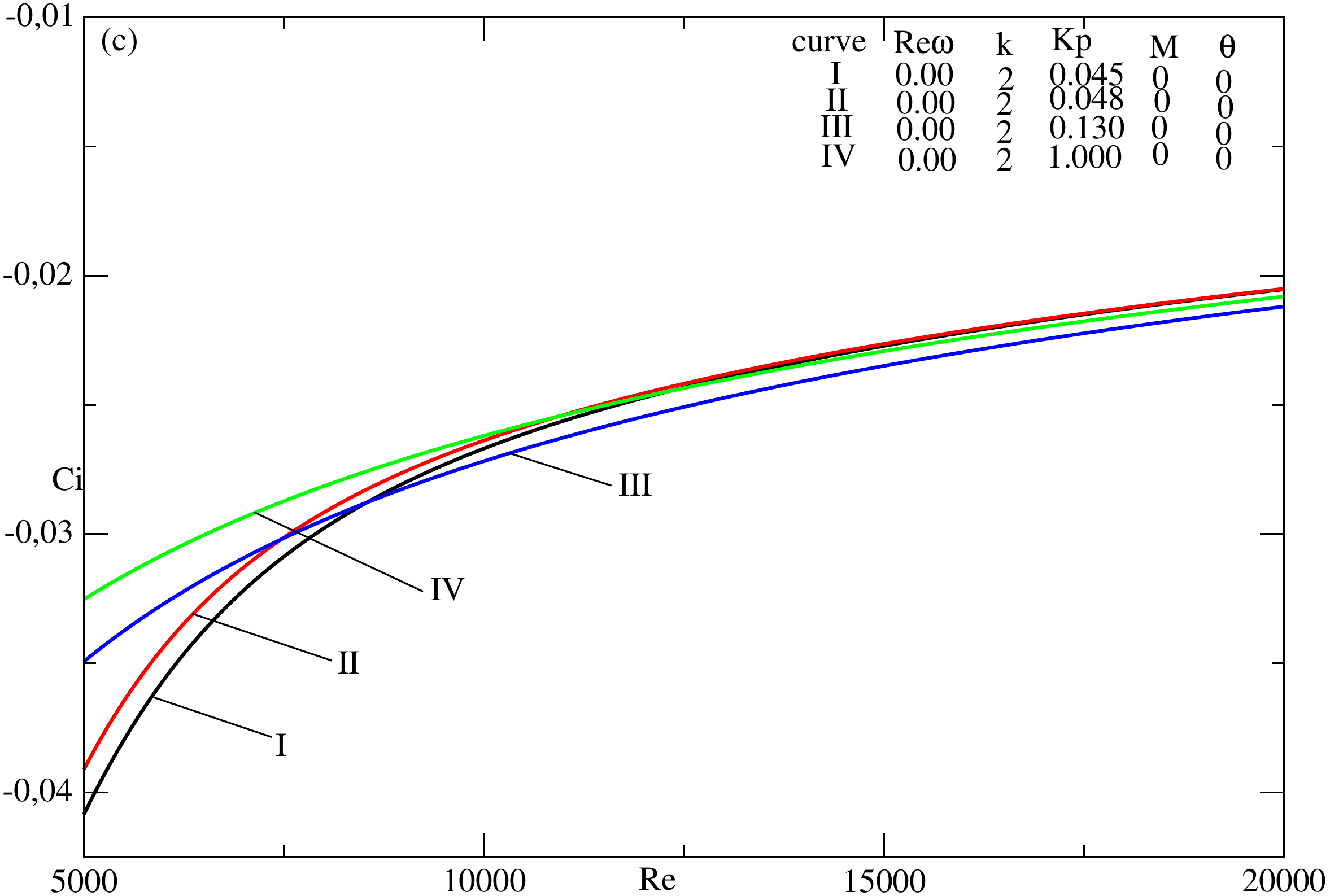} \includegraphics[width=6cm]{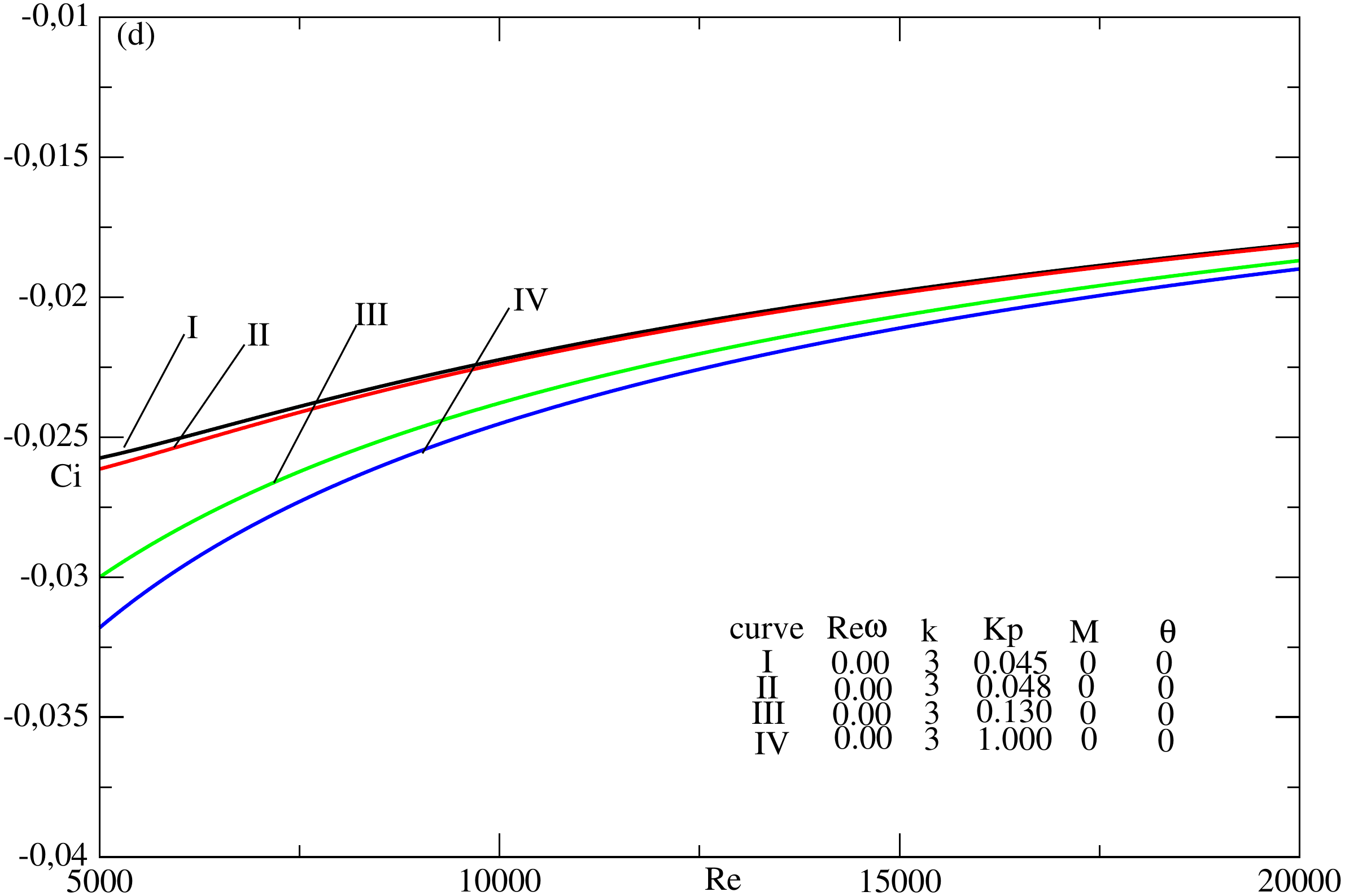}
 \end{center}
 \caption{Ci vs.Re for $R_{e\omega},  M,   \theta$ fixed and  $K_{p},  k$ variable} 
 \label{fig : 2}
 \end{figure}

 Figure \ref{fig : 2}  exhibits the effect of Permeability parameter $K_{p}$   
 on linear temporal stability of viscous incompressible  
 non electrically conduction fluid flow for  different values
 of  wave number. It is observed that  $K_{p}$  affectes the stability. 
 For $k=1$ and $k=1.02$,  the  frames $a$ and $b$ show that
 for $K_{p}=0.045$ the flow is instable (see curves $I$)
 and stable for  $ K_{p}=1.000 $ (see curves $ IV$) but for 
 $K_{p}=0.048$ and $K_{p}=0.130$  (see curves $II,  III$),  we have the transition of the flow 
 (Seen tabular below for the criticals Reynolds'number values). 
 For $k=2$ and $k=3$ (see frames $c$ and $d$) the flow is completely stable.
 On careful observation,  we remark that for $R_{e}<12500$,  $K_{p}$ 's increase
 contributes to  the stability   in frame $d$ case and the  opposite is noticed
 in the frame $c$ case,  but when $R_{e}>12500$,   $K_{p}$ 's increase contributes in the two cases. 
 Thus,  it may be concluded that except  the frame $c$ case where,  the $K_{p}$ 's increase doesn't  
 contribute to stability for $R_{e}<12500$,  the Permeability
 parameter increasing contributes  to the  flow stability.

\begin{figure}[htbp]
 \begin{center}
 \includegraphics[width=6cm]{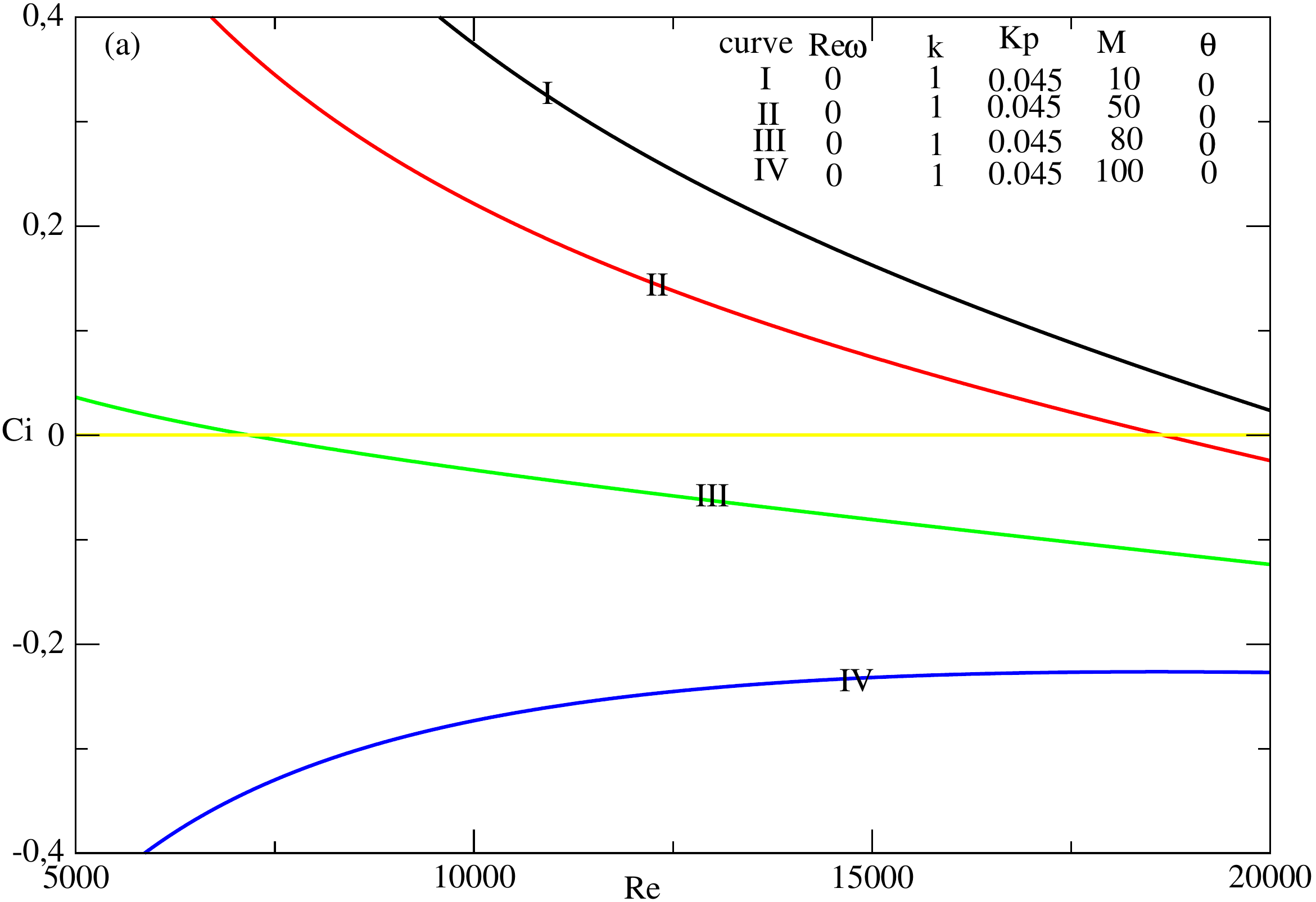} \includegraphics[width=6cm]{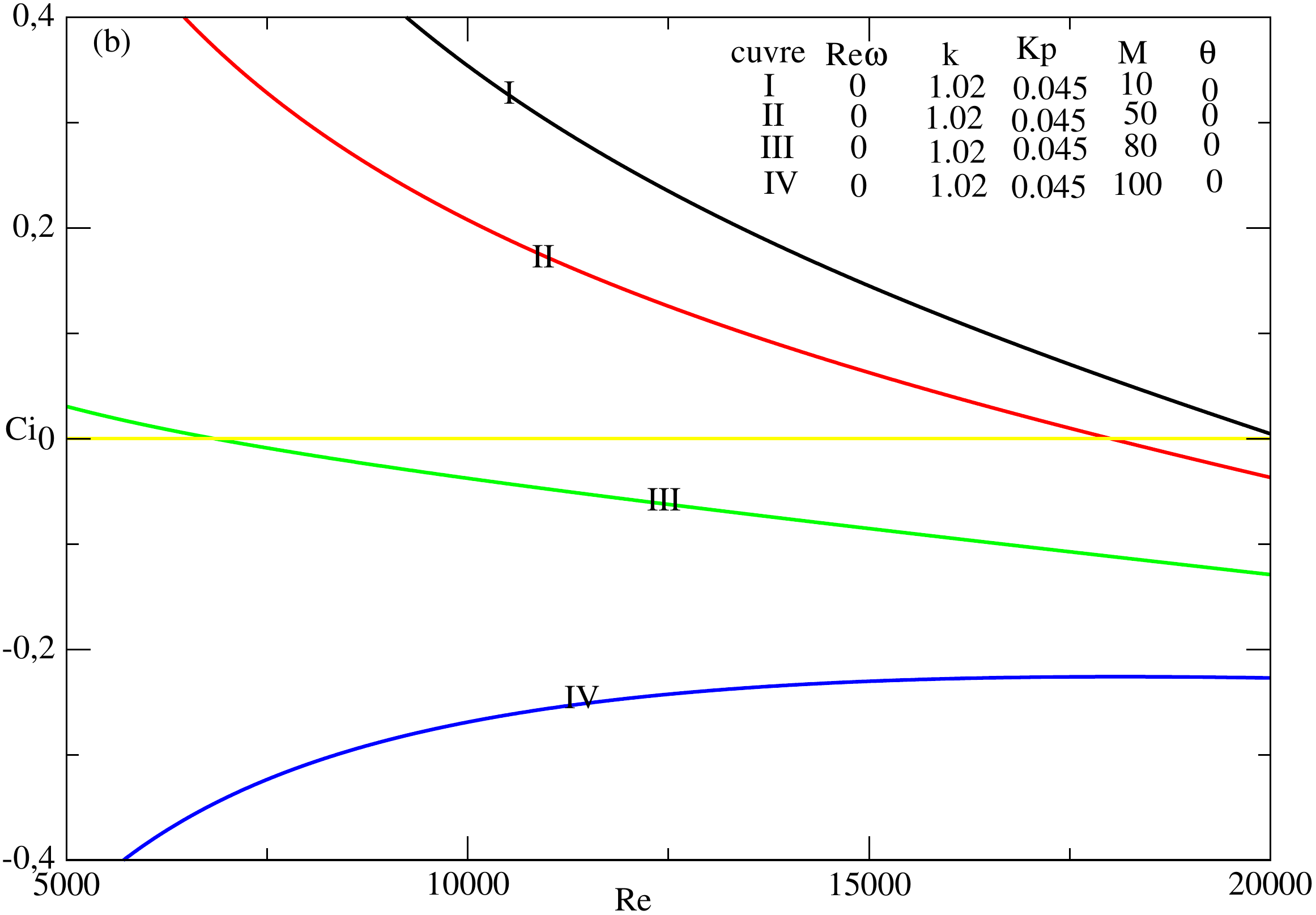} \\
 \includegraphics[width=6cm]{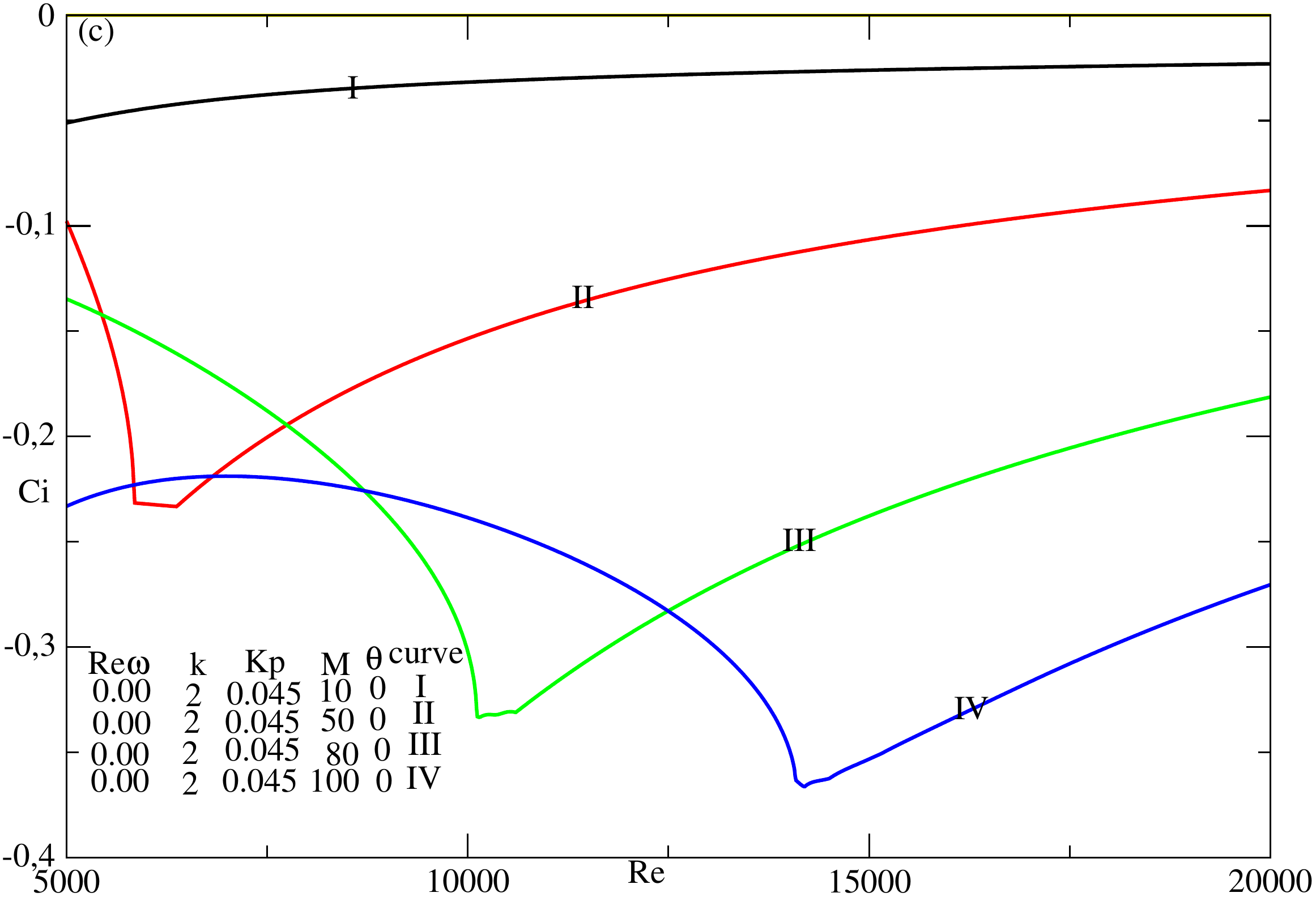} \includegraphics[width=6cm]{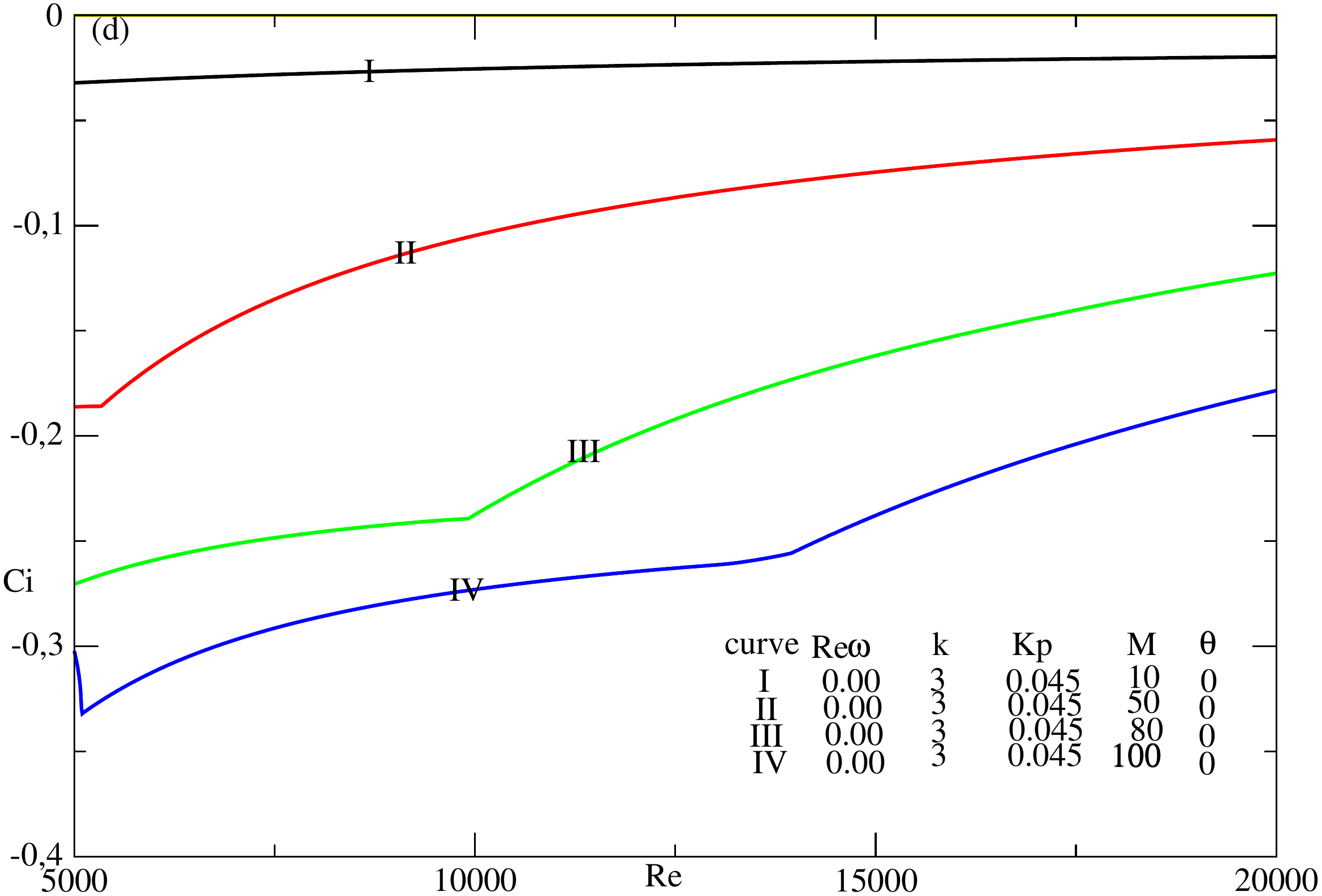}
 \end{center}
 \caption{Ci vs.Re for $R_{e\omega},  K_{p},   \theta$ fixed and  $M,  k$ variable}
 \label{fig : 3}
 \end{figure}

 Figure \ref{fig : 3}  shows the effect of Hartmann number $M$   
 on linear temporal stability of viscous incompressible  
 electrically conduction fluid   flow for  different values
 of  wave number. It is observed that  $M$  affectes the stability.
 For $k=1$ and $k=1.02$,  the  frames $a$ and $b$ show that
 for $M=10$ the flow is instable (see curves $I$) and stable for  $ M=100 $ (see curves $ IV$) but for 
 $M=50$ and $M=80$  (see curves $II,  III$),  we have the transition of the flow 
  (Seen tabular below for the criticals Reynolds'number values). 
 For $k=2$ and $k=3$ (see frames $c$ and $d$) the flow is completely stable.
 Thus,  we  may  concluded that  the Hartmann number increasing contribute  more  to the  flow stability.  
 
\begin{figure}[htbp]
 \begin{center}
 \includegraphics[width=6cm]{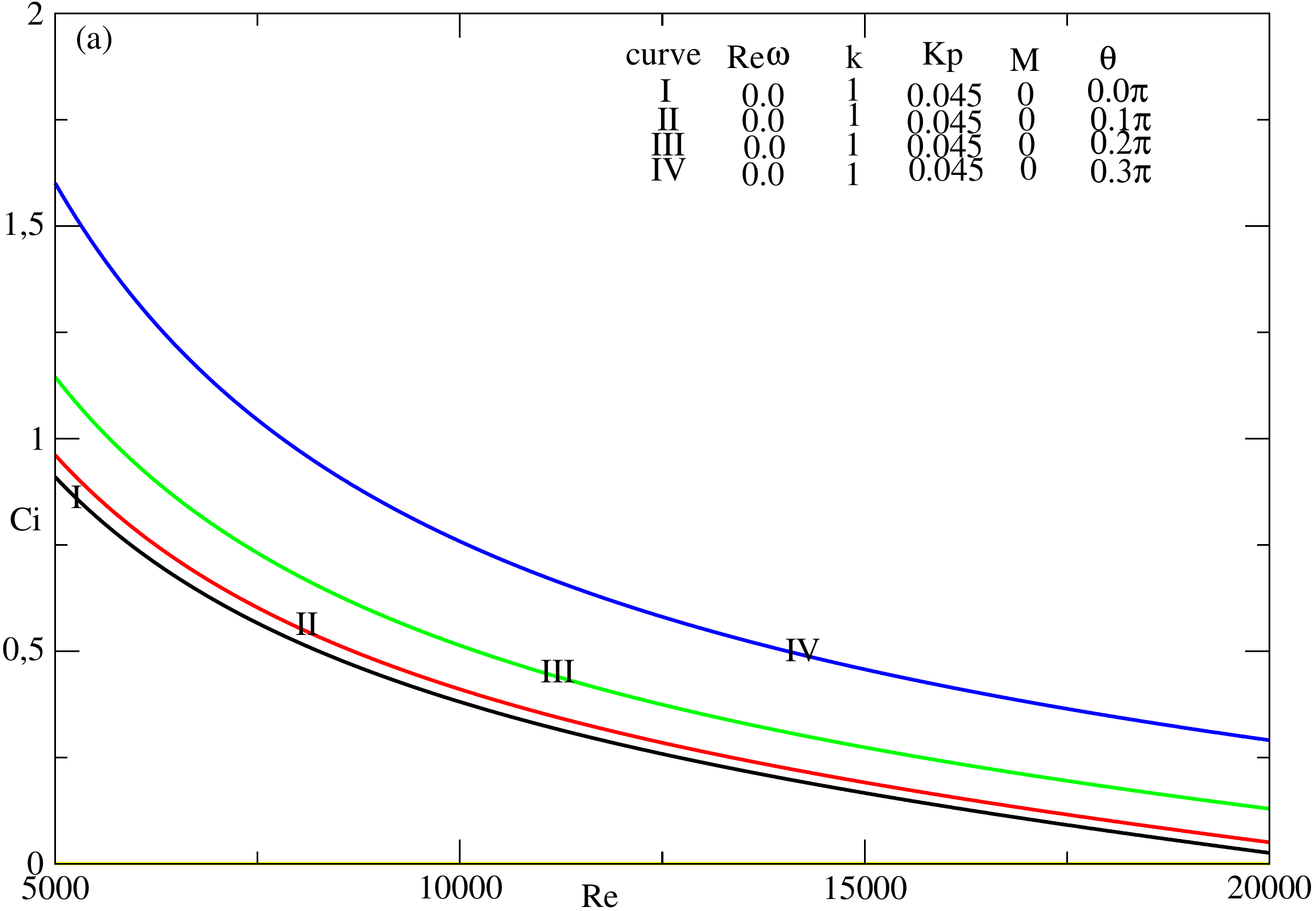} \includegraphics[width=6cm]{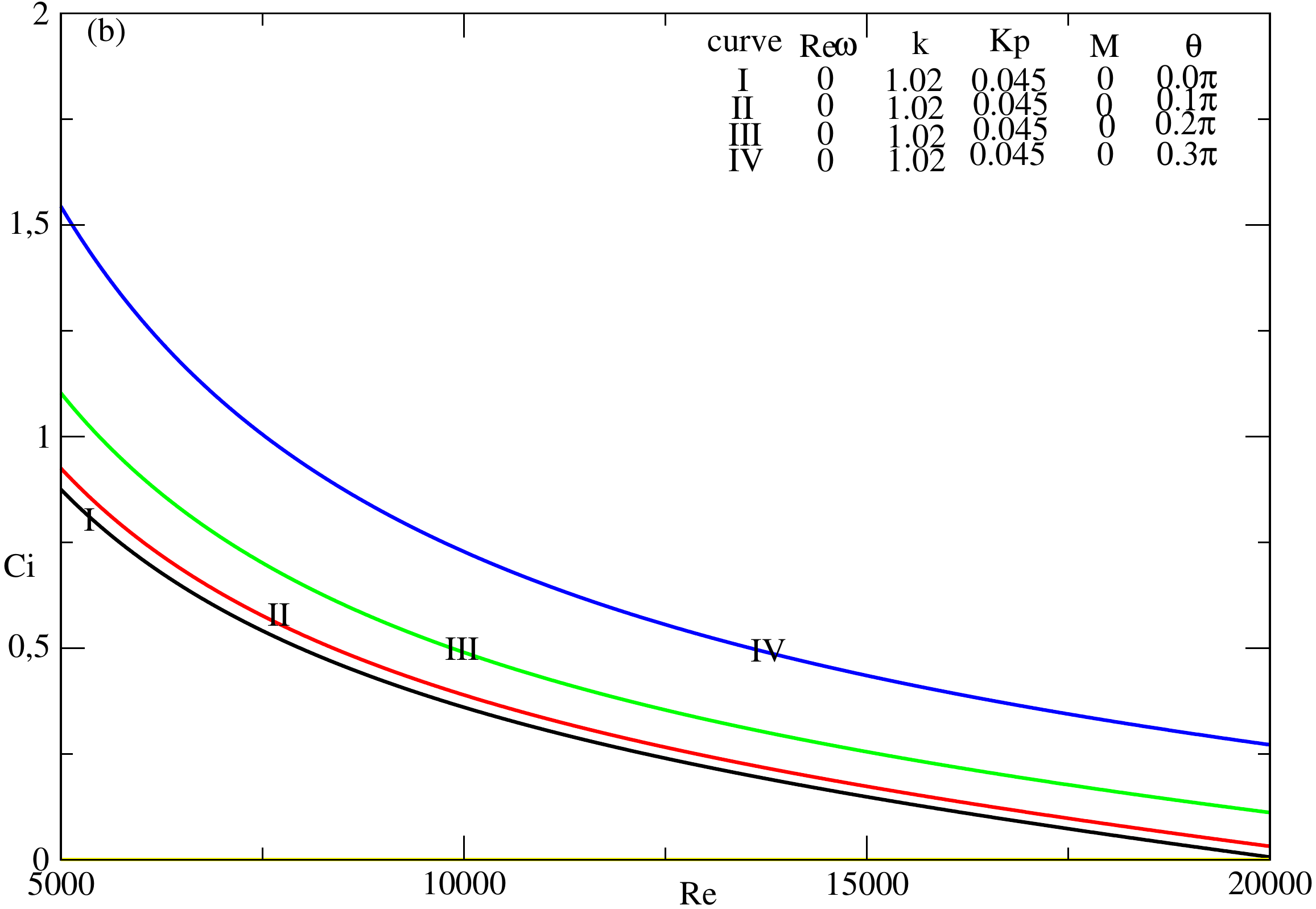} \\
 \includegraphics[width=6cm]{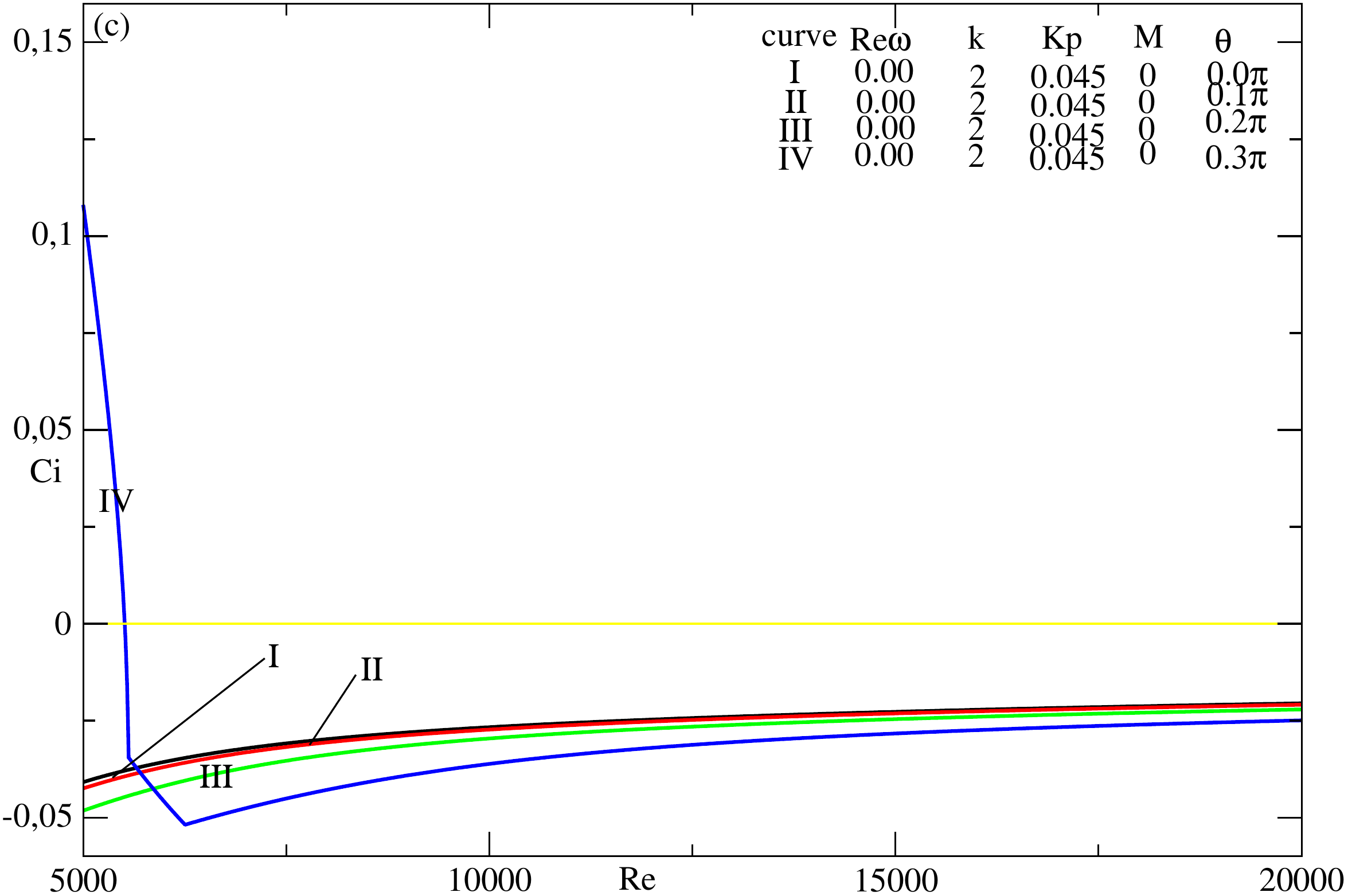} \includegraphics[width=6cm]{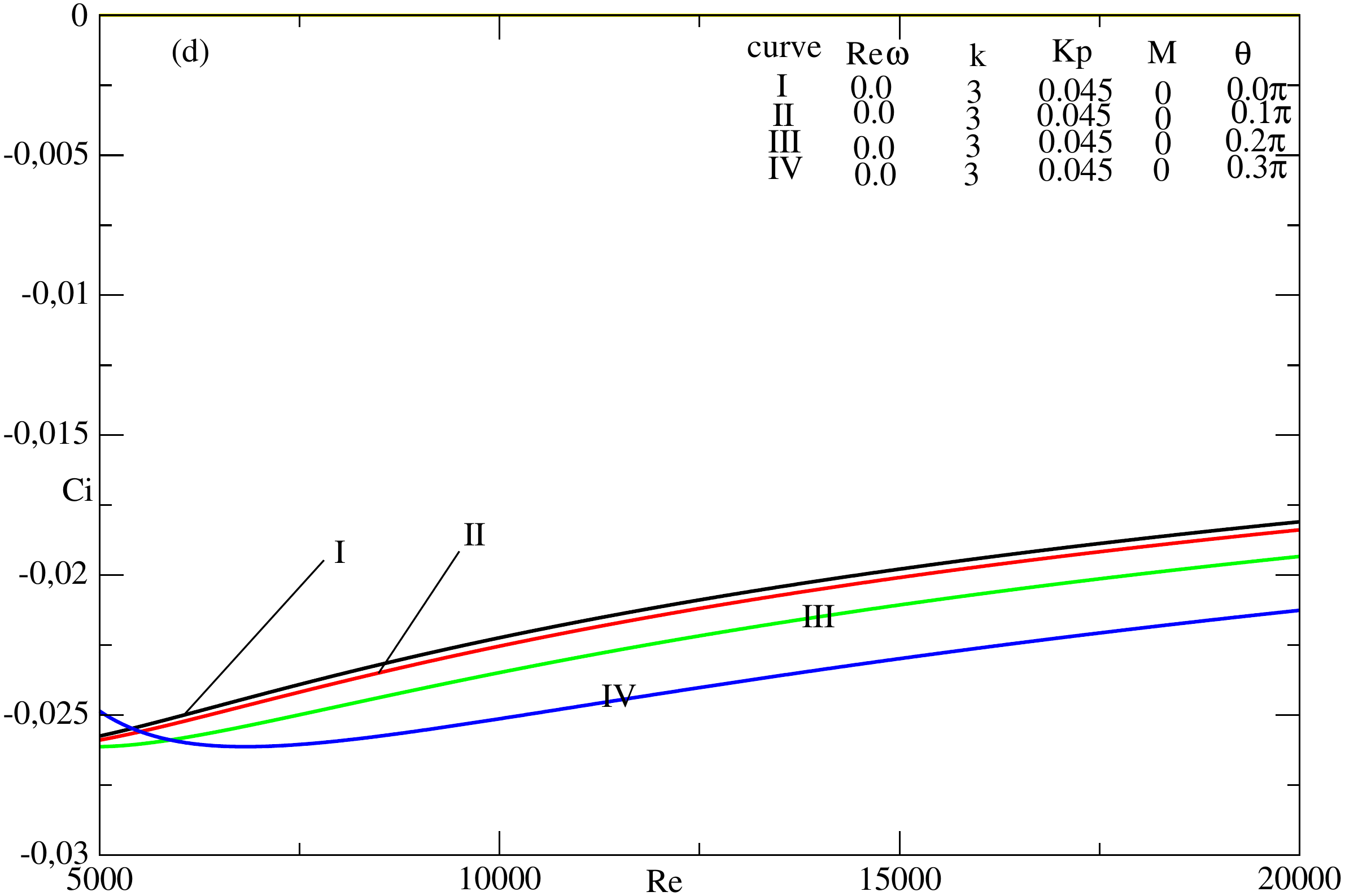}
 \end{center}
 \caption{Ci vs.Re for $R_{e\omega},  K_{p},  M$ fixed and  $ \theta,  k$ variable} 
 \label{fig : 4}
 \end{figure}

 Figure \ref{fig : 4} depicts the effet of phase angle $\theta$
 for different values of the wave number on the flow stability.
 For $k=1$ and $k=1.02$,  the  frames $a$ and $b$ show that the flow is instable and the 
 instability increases when the  angle $\theta$ increases. But,  for  $k=2$ and $k=3$  
 the flow is completely stable except the curve $IV$ frame $c)$,  
 which presents a transition initialy  and stays instable  after.  
 
\begin{figure}[htbp]
 \begin{center}
 \includegraphics[width=6cm]{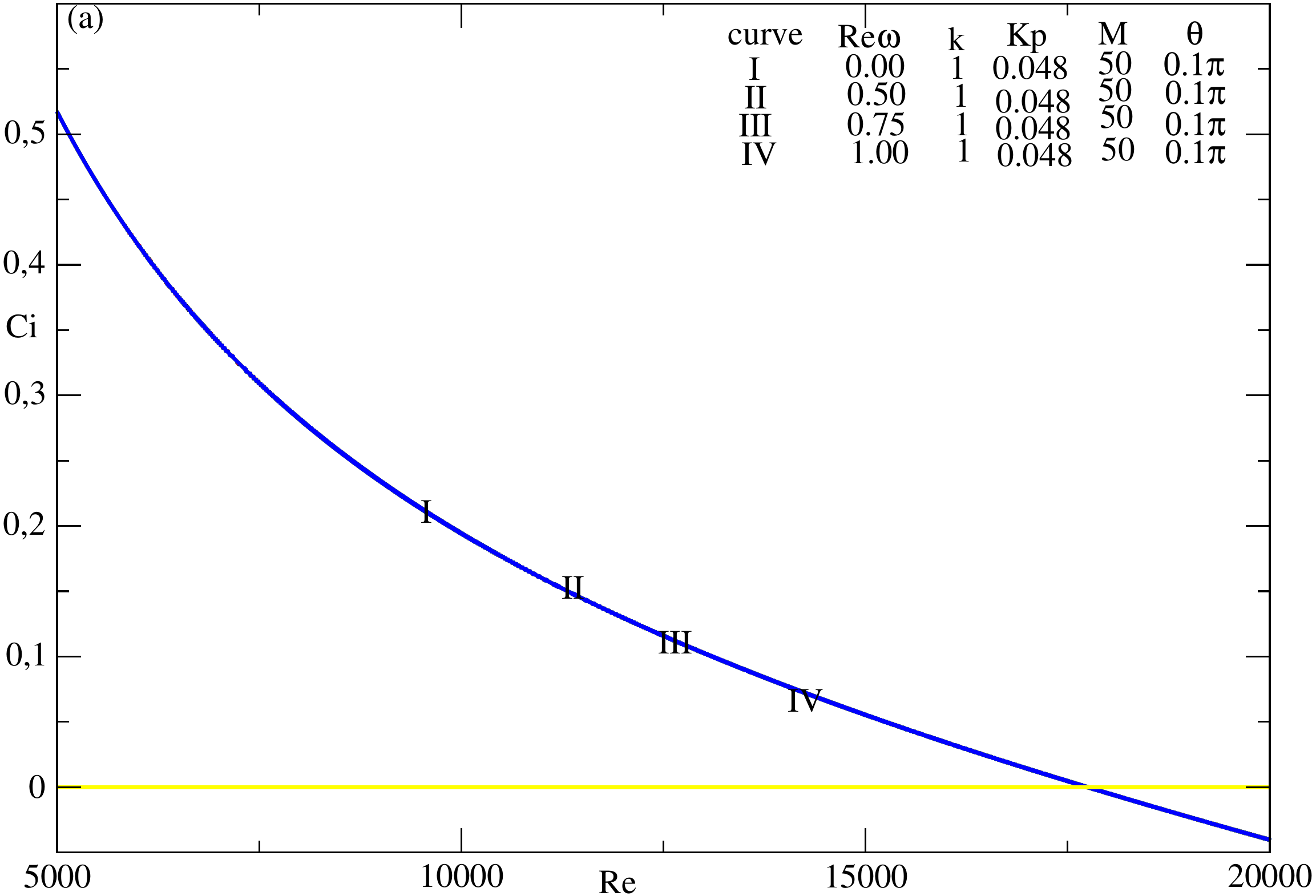} \includegraphics[width=6cm]{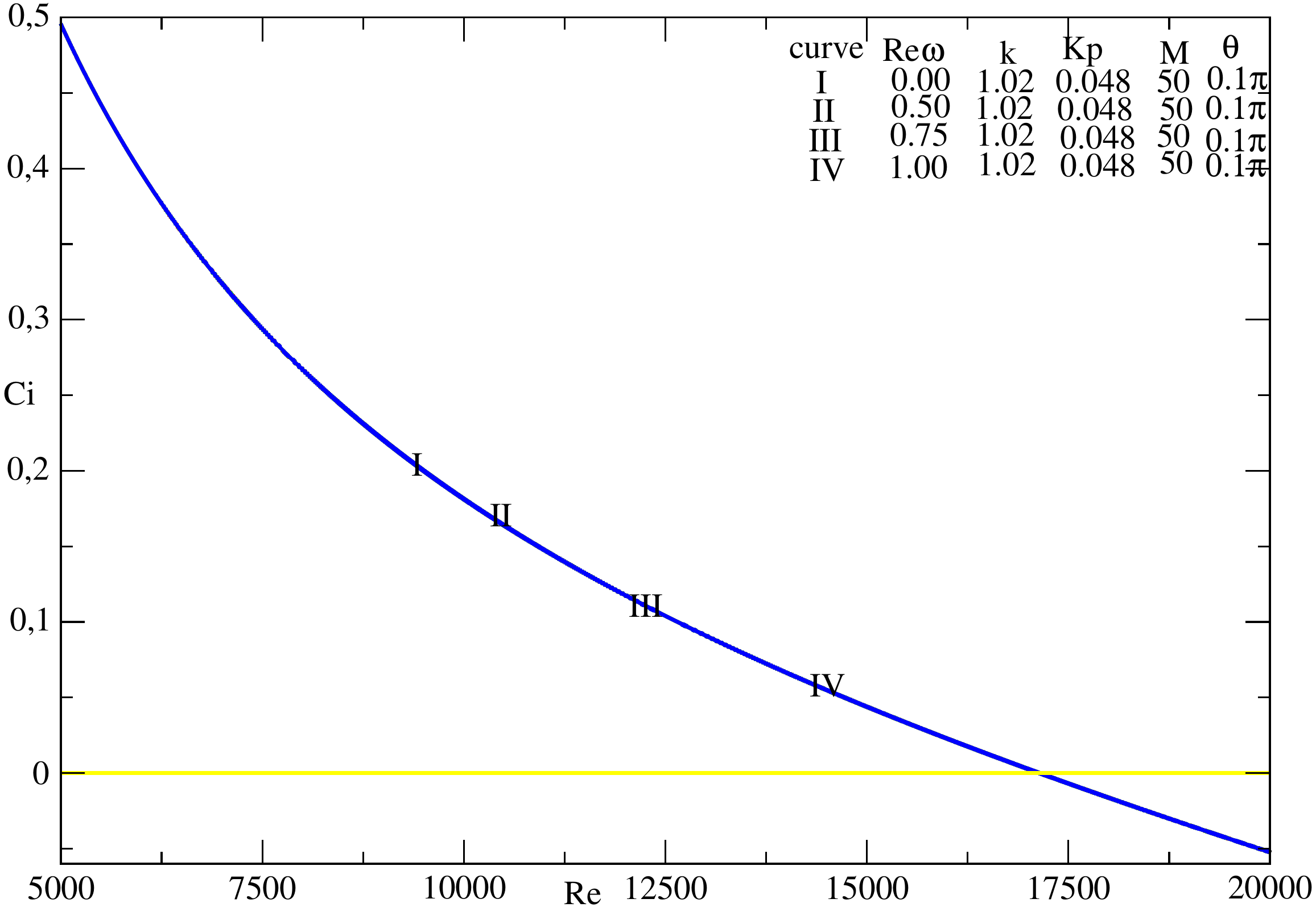} \\
 \includegraphics[width=6cm]{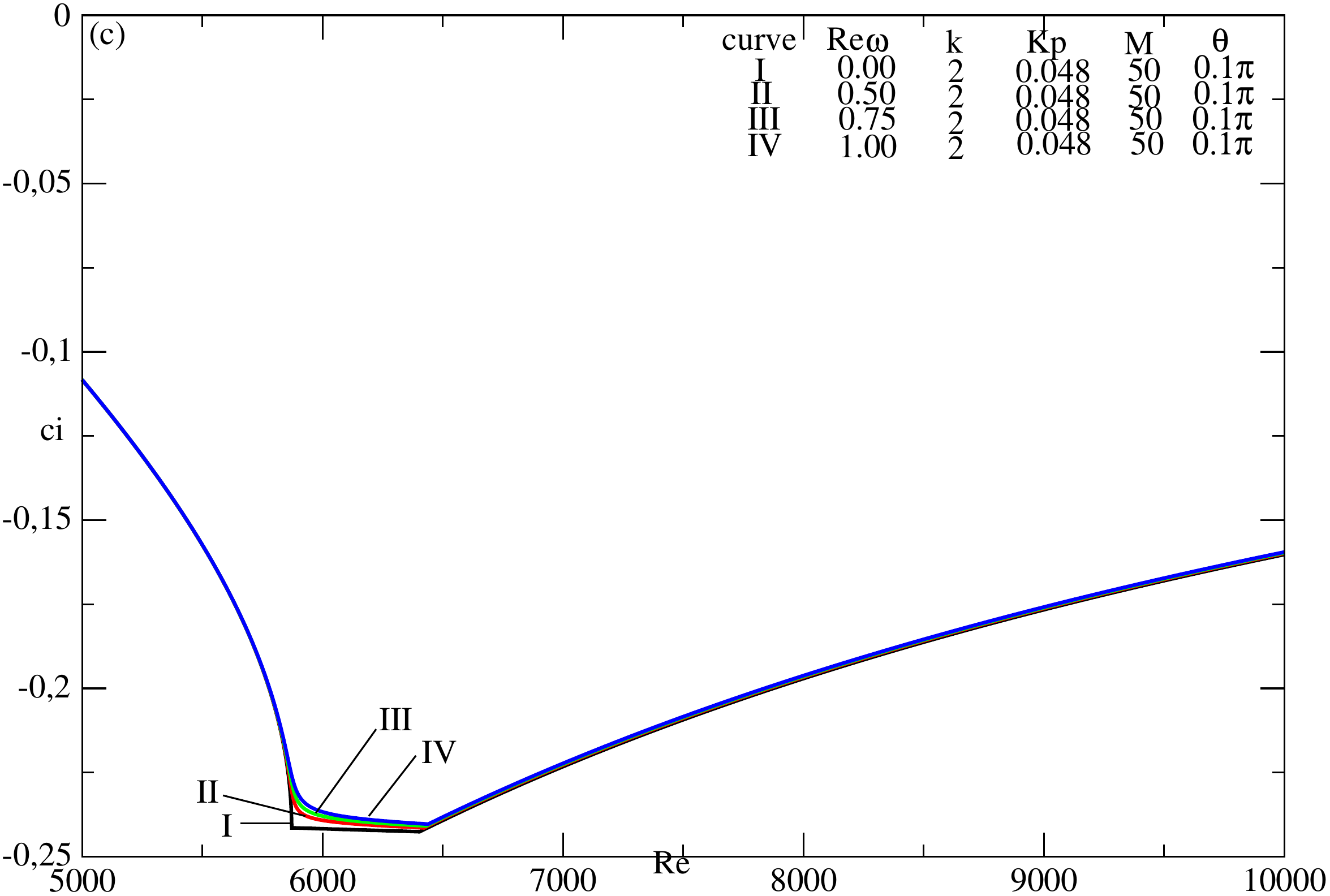} \includegraphics[width=6cm]{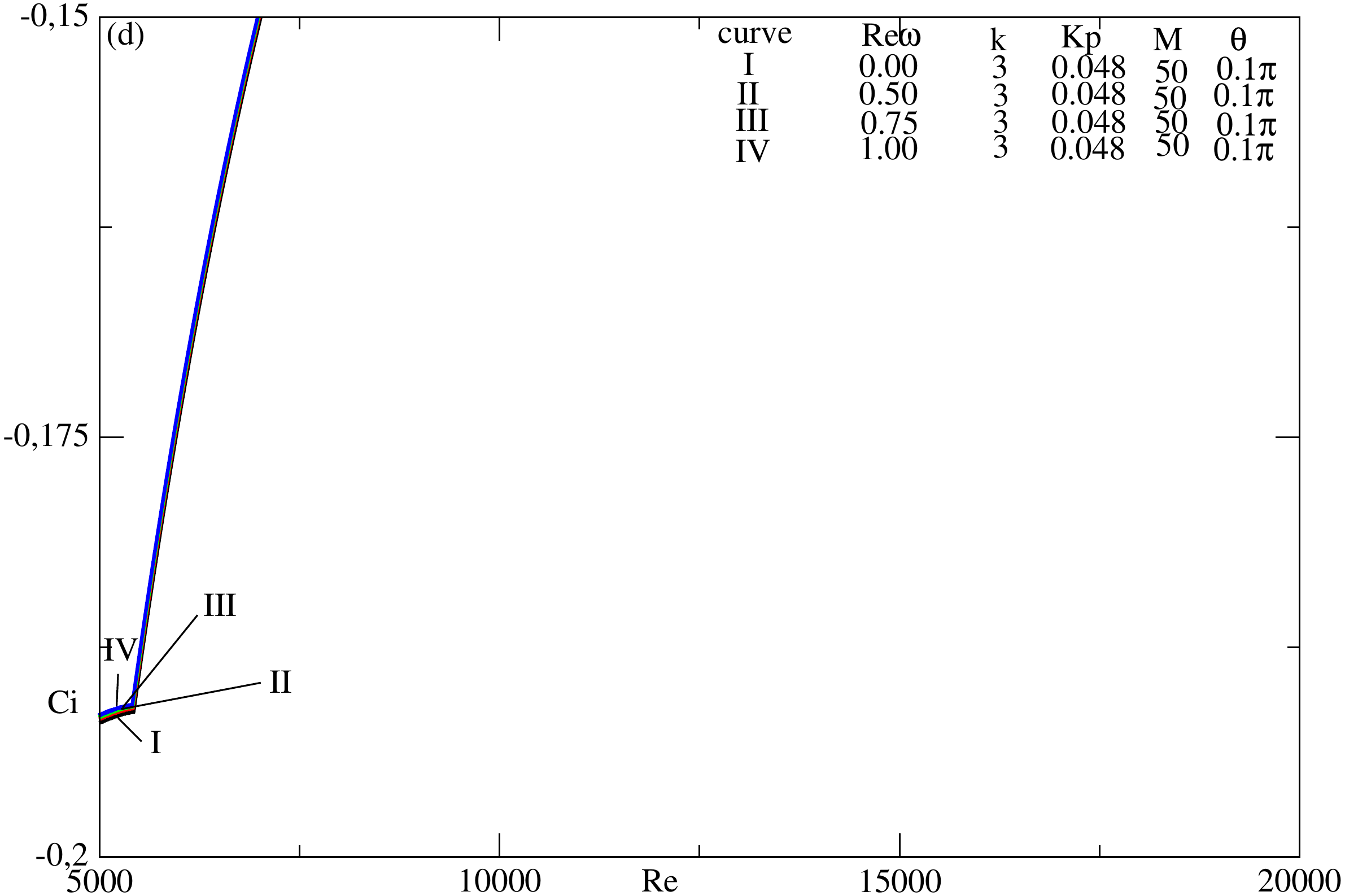}
 \end{center}
 \caption{Ci vs.Re for $\theta ,  M$ fixed and  $  R_{e\omega},  K_{p},  k$ variable} 
 \label{fig : 5}
 \end{figure}

\begin{figure}[htbp]
 \begin{center}
 \includegraphics[width=6cm]{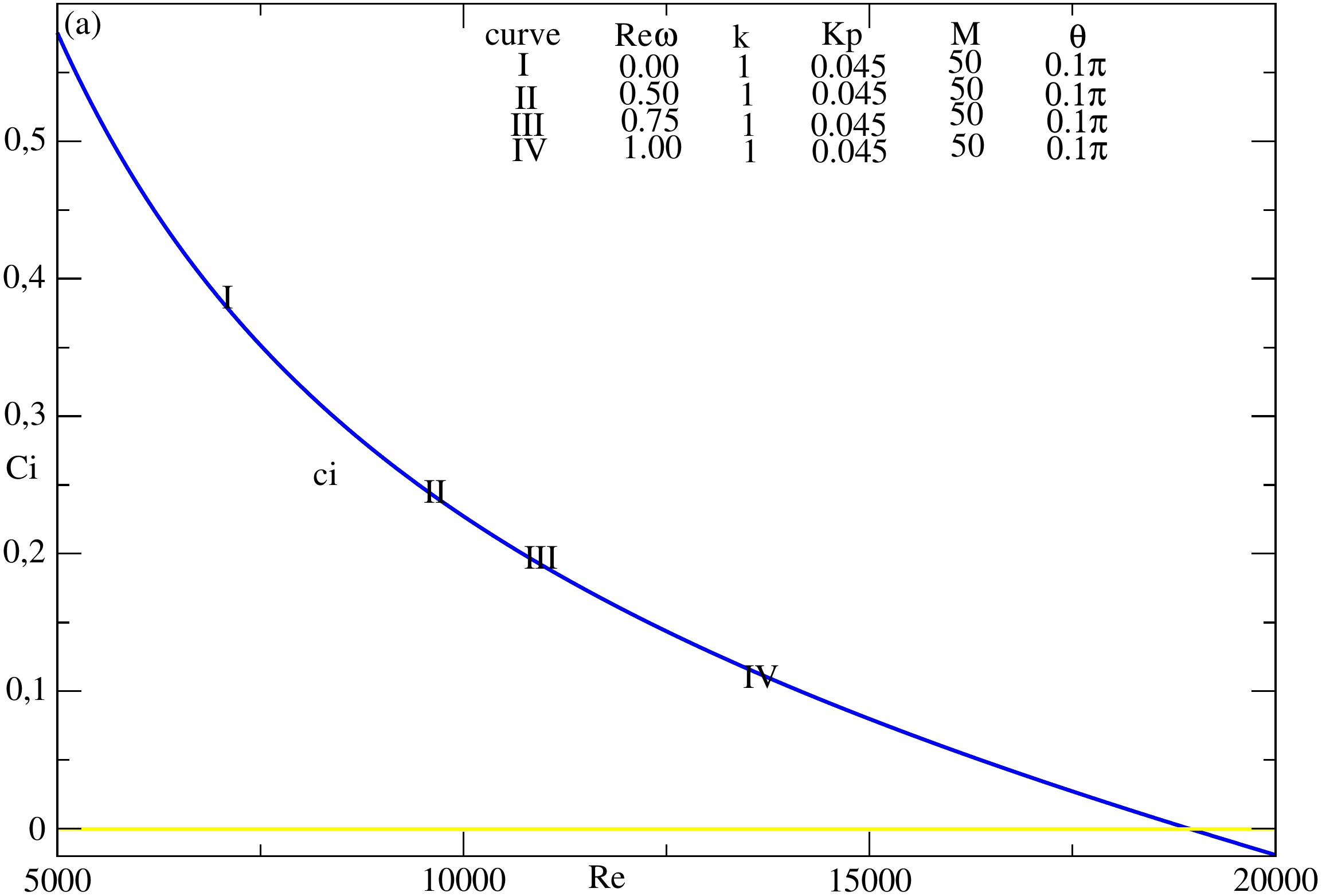} \includegraphics[width=6cm]{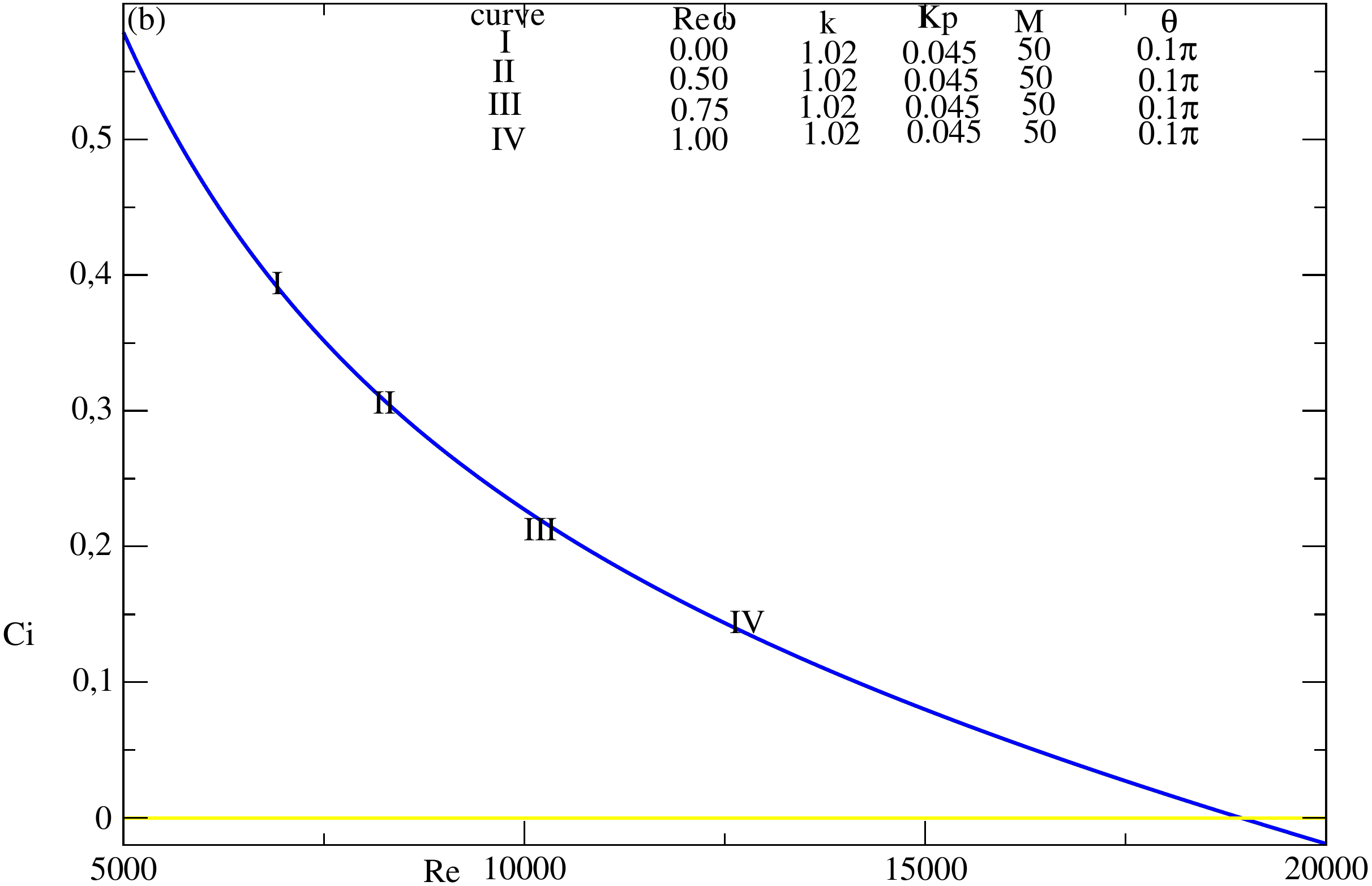} \\
 \includegraphics[width=6cm]{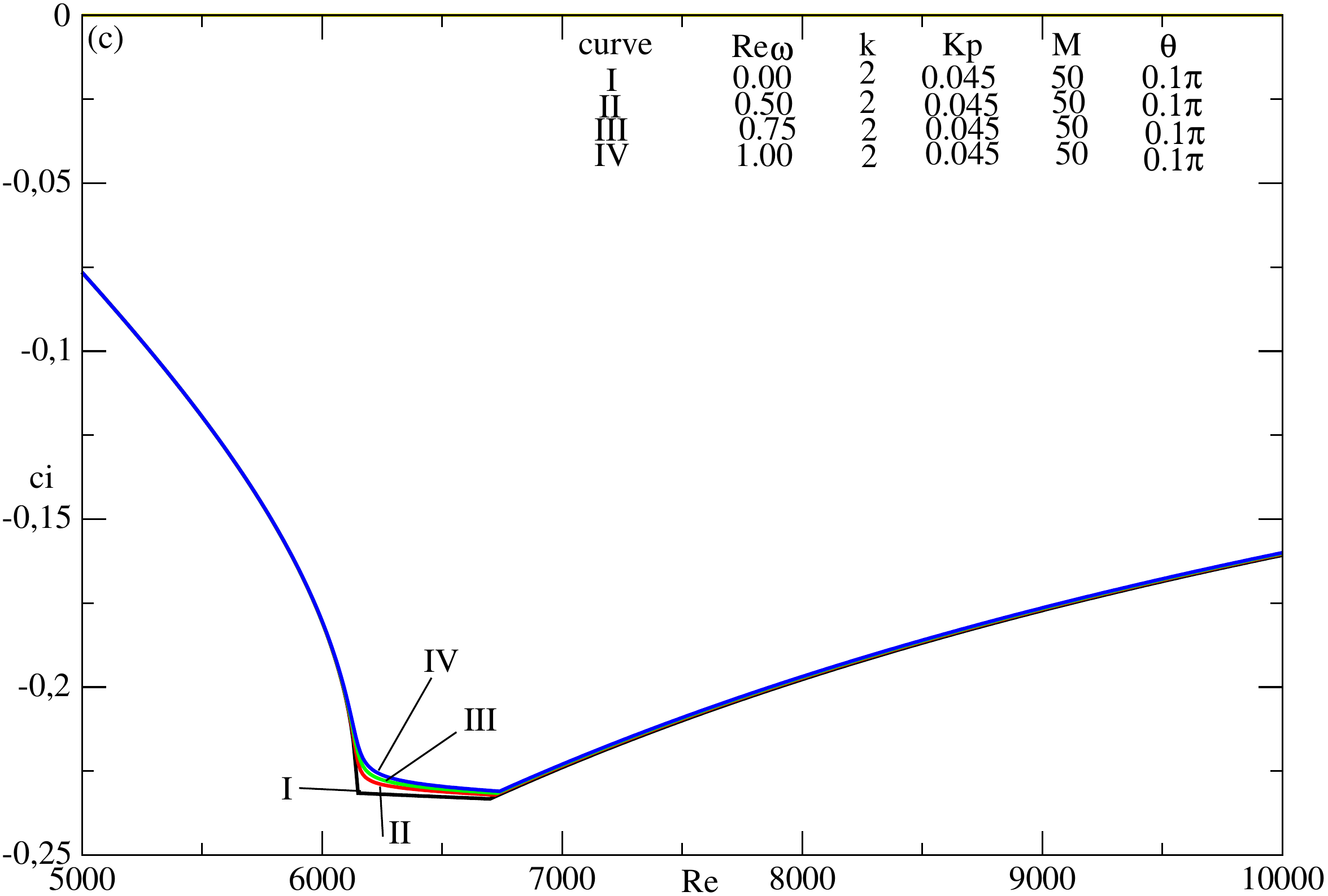} \includegraphics[width=6cm]{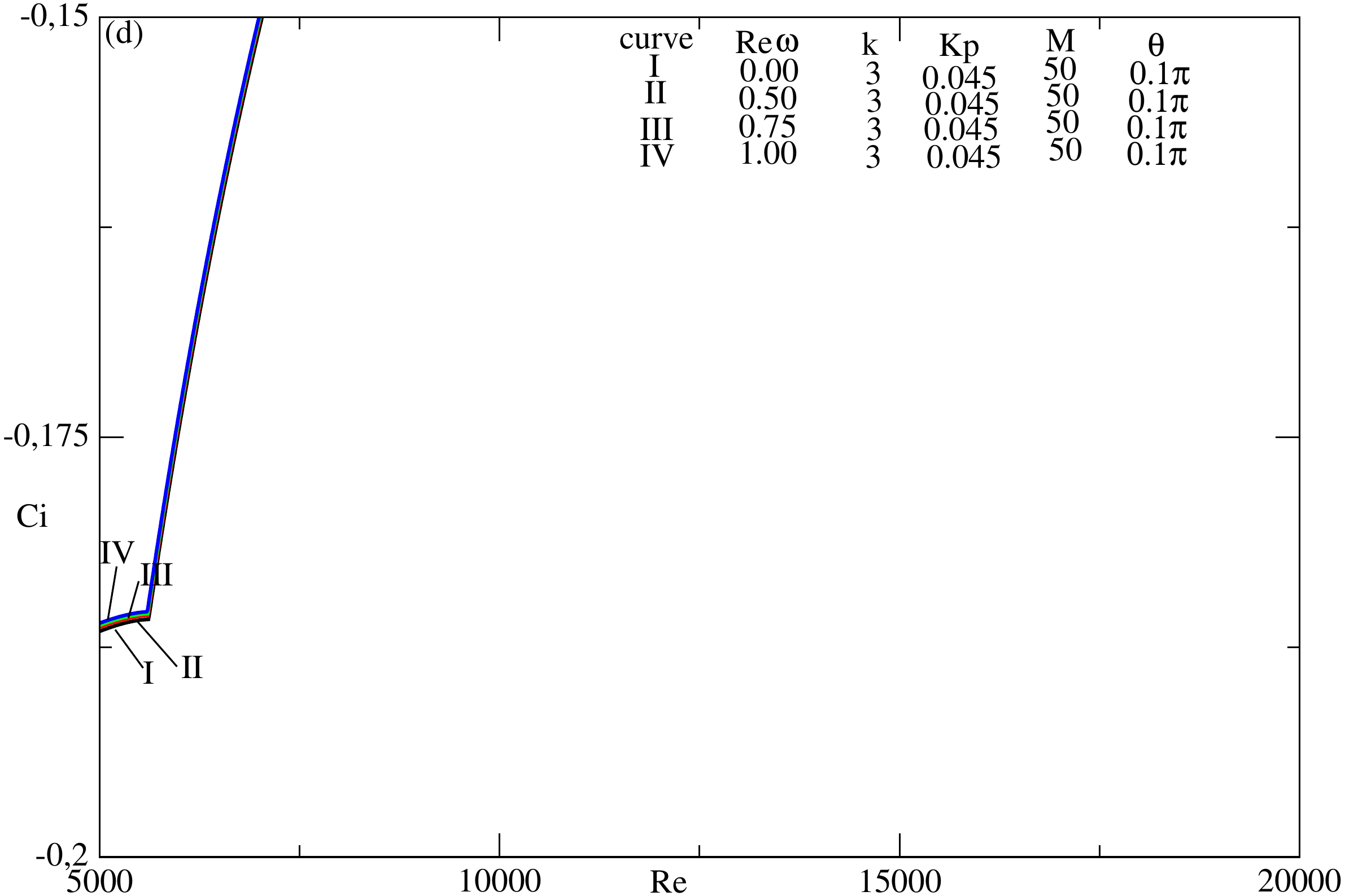}
 \end{center}
 \caption{Ci vs.Re for $\theta,  K_{p},   M$ fixed and  $R_{e\omega},  k$ variable} 
 \label{fig : 6}
 \end{figure}

\begin{figure}[htbp]
 \begin{center}
 \includegraphics[width=6cm]{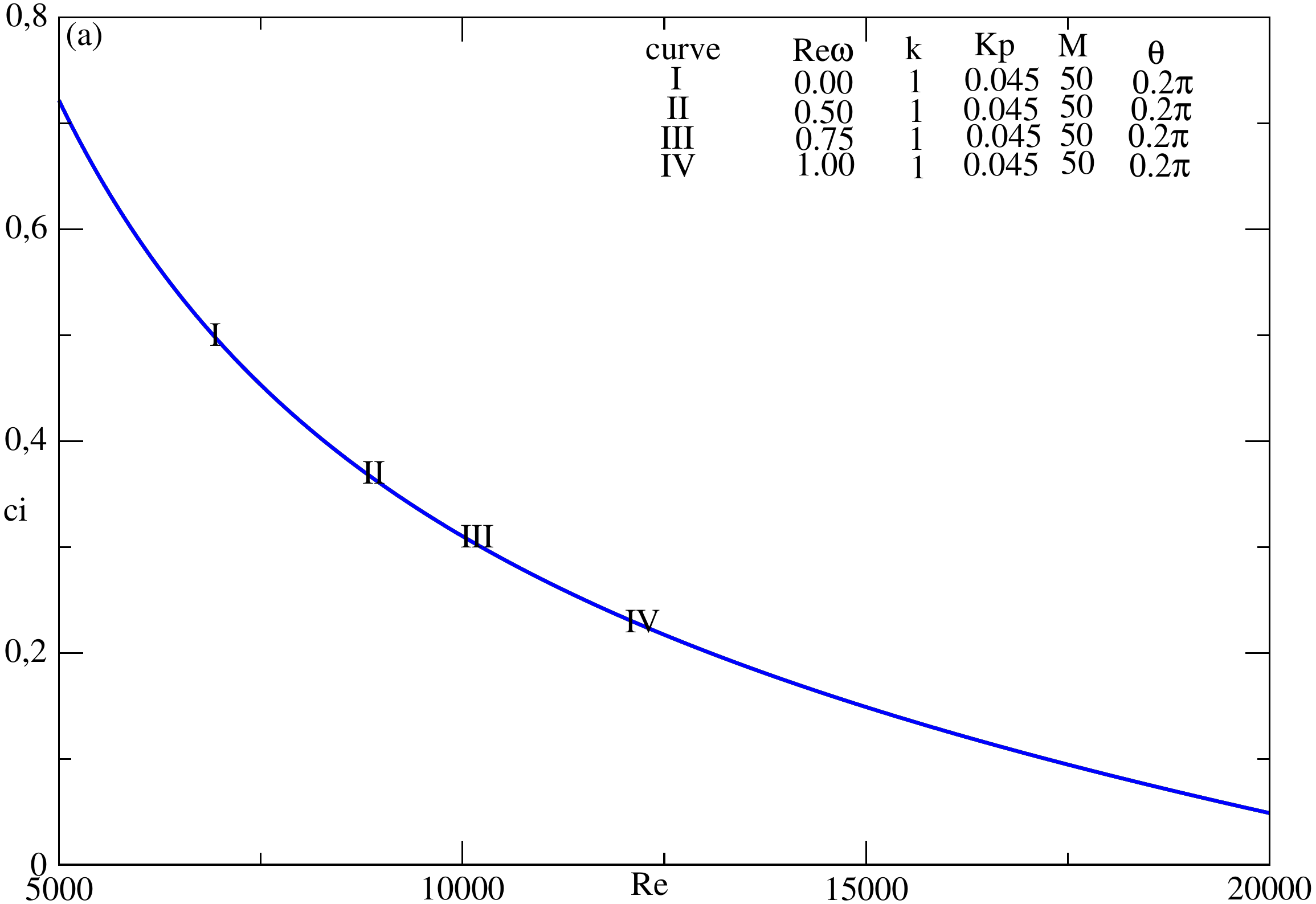} \includegraphics[width=6cm]{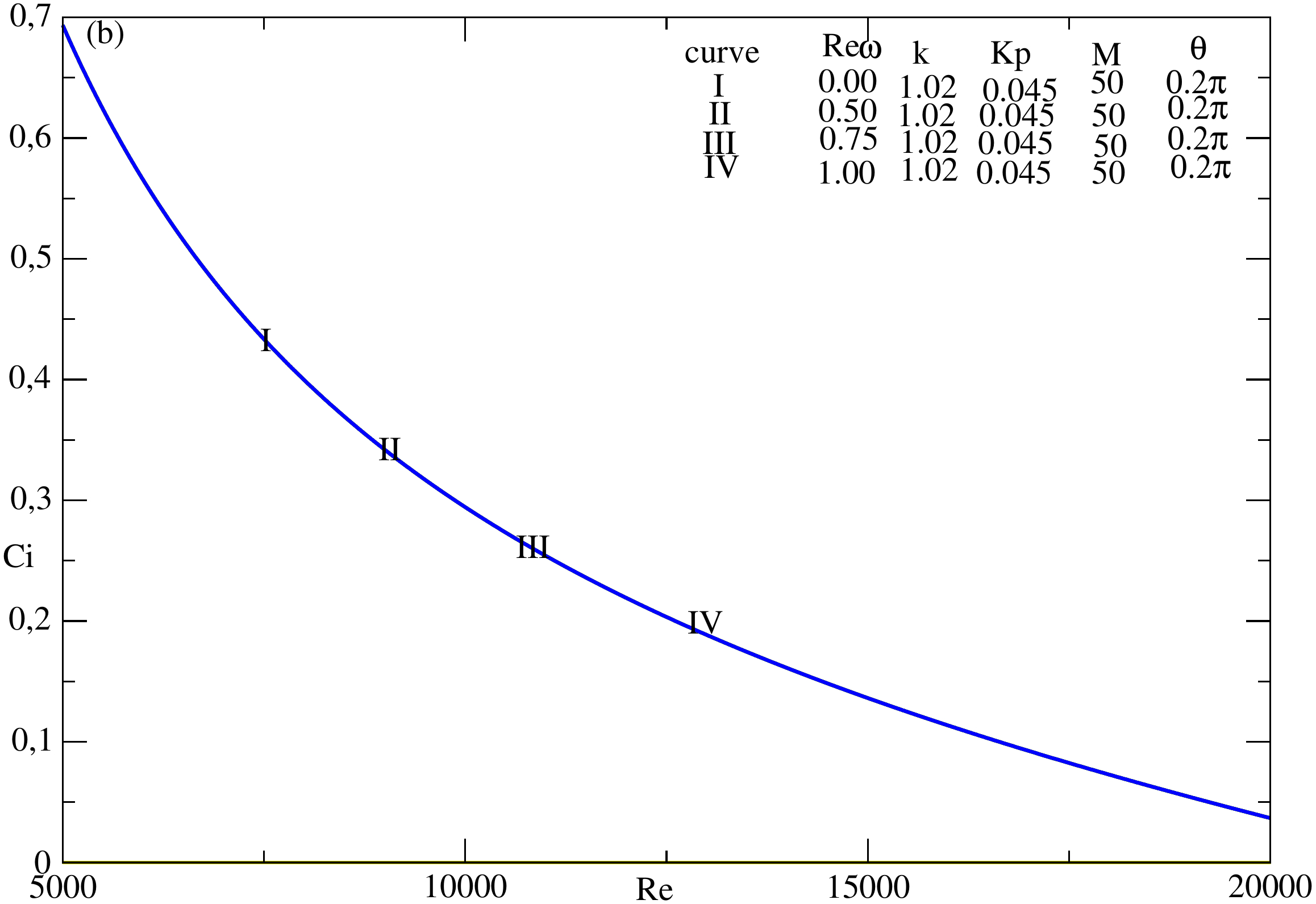} \\
 \includegraphics[width=6cm]{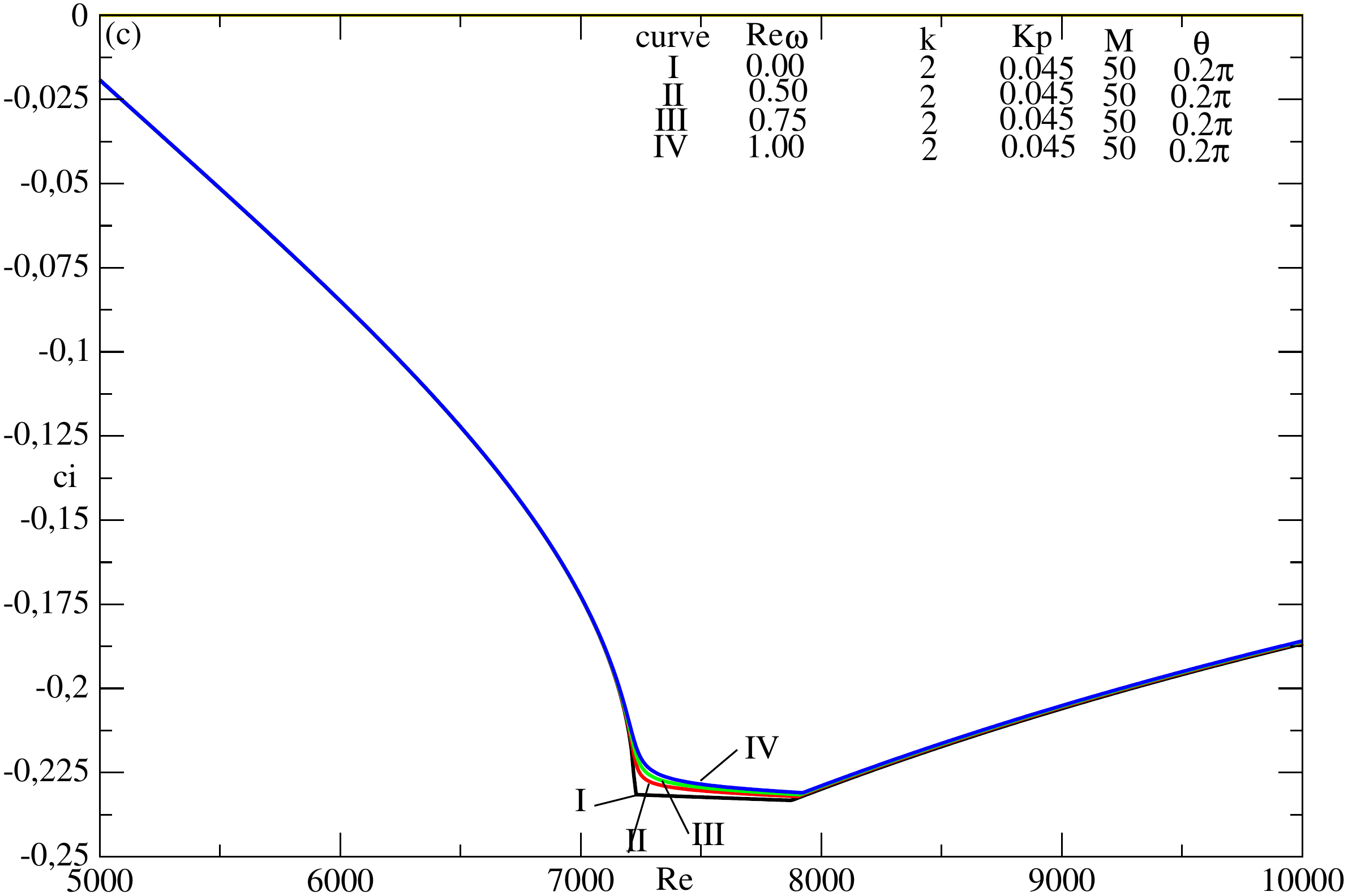} \includegraphics[width=6cm]{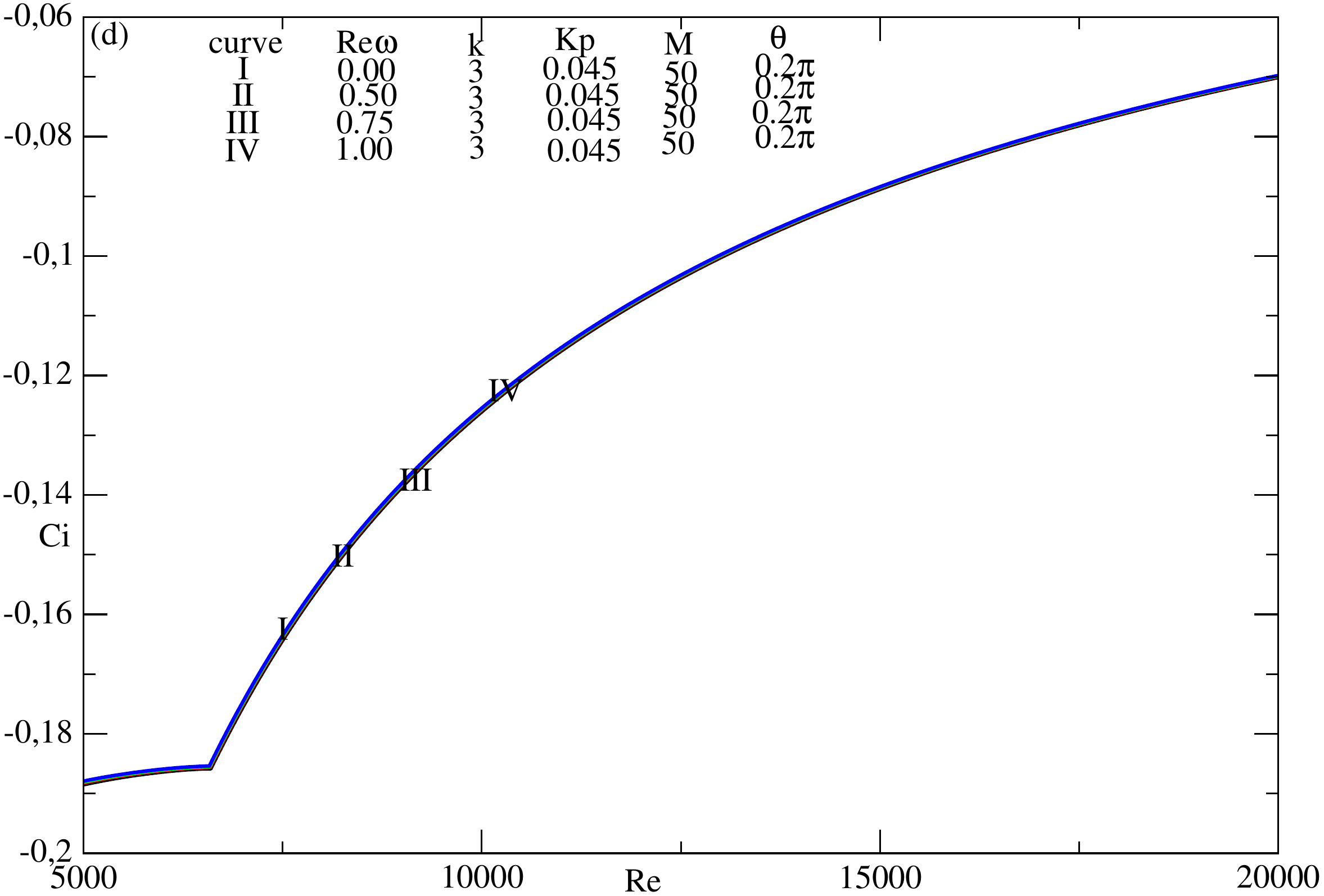}
 \end{center}
 \caption{Ci vs.Re for $\theta , K_{p},   M$ fixed and  $  R_{e\omega},  k$ variable} 
 \label{fig : 7}
 \end{figure}

 Finaly,  the Figures \ref{fig : 5},  \ref{fig : 6} and \ref{fig : 7}  
 ( $M\neq0$,  electically conduction fluid) show that for $k=1$ and $1.02$ the small injection/suction has no   effect on 
 the linear temporal stability of the flow. But for $k=2$ and $3$,  we remark  a little influence of the small injection/suction
 on the stability only in  a little interval of $R_{e}$.

 \begin{center}
\begin{tabular}{|l|l|l|l|l|l|}
\hline
$R_{e\omega}$ & $k$ & $Kp$ &$M$ &$\theta (\pi)$ & $R_{ec}$\\
\hline
 $0.00$&$1.00$ & $0.130$ & $00$ & $0.00 $ &$05347$\\
 \hline
$0.00$ &$1.00$ &$0.048$ & $00$ &$0.00$ & $19370$\\
\hline
$0.00$ &$1.02$ & $0.130$ &$00$ &$0.00$ &$05139$\\
\hline
$0.00$ &$1.02$ & $0.048$ &$00$ &$0.00$ &$18640$\\
\hline
$0.00$ & $1.00$ & $0.045$ &$50$ &$0.00$ &$18650$\\
\hline
$0.00$ & $1.00$ & $0.045$ &$80$ &$0.00$ &$07178$\\
\hline
$0.00$ & $1.02$ & $0.045$ &$50$ &$0.00$ &$18010$\\
\hline
$0.00$ & $1.02$ & $0.045$ &$80$ &$0.00$ &$06845$\\

\hline

$0.00$ & $1.00$ & $0.045$ &$50$ &$0.10$ &$19610$\\
\hline
$0.50$ & $1.00$ & $0.045$ &$50$ &$0.10$ &$19610$\\
\hline
$0.75$ & $1.00$ & $0.045$ &$50$ &$0.10$ &$19610$\\
\hline
$1.00$ & $1.00$ & $0.045$ &$50$ &$0.10$ &$19610$\\
\hline
$0.00$ & $1.02$ & $0.045$ &$50$ &$0.10$ &$18940$\\
\hline
$0.50$ & $1.02$ & $0.045$ &$50$ &$0.10$ &$18940$\\
\hline
$0.75$ & $1.02$ & $0.045$ &$50$ &$0.10$ &$18940$\\
\hline
$1.00$ & $1.02$ & $0.045$ &$50$ &$0.10$ &$18940$\\

\hline
$0.00$ & $1.00$ & $0.048$ &$50$ &$0.10$ &$17750$\\
\hline
$0.50$ & $1.00$ & $0.048$ &$50$ &$0.10$ &$17750$\\
\hline
$0.75$ & $1.00$ & $0.048$ &$50$ &$0.10$ &$17750$\\
\hline
$1.00$ & $1.00$ & $0.048$ &$50$ &$0.10$ &$17750$\\
\hline
$0.00$ & $1.02$ & $0.048$ &$50$ &$0.10$ &$17140$\\
\hline
$0.50$ & $1.02$ & $0.048$ &$50$ &$0.10$ &$17140$\\
\hline
$0.75$ & $1.02$ & $0.048$ &$50$ &$0.10$ &$17140$\\
\hline
$1.00$ & $1.02$ & $0.048$ &$50$ &$0.10$ &$17140$\\
\hline

 \end{tabular}
  \end{center}

 \newpage
 \section*{figures}

 \newpage
 
 \section*{Conclusion}

 The conclusions of the study are : 
 \begin{itemize}
  \item 
 The small injection/suction Reynolds' number has  a little effect on the linear temporal stability of
the  couette flow only for the  high waves numbers ($k\geq2 $).
\item  The  permeability parameter i.e. Darcy number,  the Lorentz force parameter i.e. the Hartmann number and 
 the wave number contribute to the linear temporal stability of the  couette flow.
 \item  For the small wave number $(k\simeq 1)$,  $\theta$ doesn't contribute to the stability of the couette flow but for $k=2,  3$  his effect is opposite. 
 \end{itemize}


\begin{thebibliography}{100}
 \bibitem{1} Fluid Mechanics,  $SG2214$,  $HT2010$ September $14$,  $2010$.
 \bibitem{2}A. V. Monwanou  and J. B. Chabi Orou., 
 \emph{The Inviscid Instability in an Electrically Conducting Fluid Affected by a Parallel Magnetic Field }
(The African Review of Physics ($2012$) $7 : 0044$).

\bibitem{3}K. Chand,  R. Kumar and S. Sharma ., 
\emph {Hydromagnetic Oscillatory Flow through a Porous Medium Bounded by two
Vertical Porous Plates with Heat Source and Soret Effect}
(Advances in Applied Science Research,  $2012$,  $3 (4) : 2169-2178$).

\bibitem{4}A. NAYAK,  G.C. DASH., 
\emph {Oscillatory effect on magneto-hydrodynamic
flow and heat transfer in a rotating horizontal porous channel} 
(ANNALS OF FACULTY ENGINEERING HUNEDOARA-International Journal Of Engineering :  p$199-208$-Tome $XI$ (Year $2013$).
Fascicule $1$. ISSN $1584 - 2665$).

\bibitem{5}N. V. R. V. Prasad,  G. S. S. Raju,  S. Venkataraman., 
\emph{Unsteaedy Hydromagnetic Flow Through A Porous Medium In
A Horizontal Channel Under Prescribed Discharge With Inclined Magnetic Field}
(International Journal of Emerging Technology and Advanced Engineering
ISSN $2250-2459$,  Volume $2$,  Issue $7$,  July $2012$).

\bibitem{6}S. S. Das,  M. Mohanty,  S. K. Panigrahi,  R. K. Padhy,  M. Sahu.,  
\emph{Radiative Heat and Mass Transfer Effects on Natural Convection Couette Flow 
through a Porous Medium in the Slip flow Regime}
(International Journal of Renewable Energy Technology Research
Vol. $1$,  No.$1$,  PP :  $01- 14$,  December $2012$,  ISSN :  $2325-3924$ (Online)).

\bibitem{7}Shalini,  Prof. (Dr.) M. S. Saroa,  Prof. (Dr.) Rajeev Jha., 
\emph{Unsteady Flow of a Dusty Conducting Fluid through porous medium
between Parallel Porous Plates with Temperature Dependent Viscosity and Heat Source}
(International Journal of Scientific Research Engineering and Technology (IJSRET)
Volume $1$ Issue$3$ pp $013-021$ June $2012$).
\bibitem{8}M. E. Sayed-Ahmed,  Hazem A. Attia,  Karem M. Ewis., 
\emph{Time Dependent Pressure Gradient Effect on Unsteady
MHD Couette Flow and Heat Transfer of a Casson Fluid}
(Engineering,  $2011$,  $3$,  $38$-$49$
doi : $10.4236$/eng.$2011.31005$ Published Online January $2011$ (http : //www.scirp.org/journal/eng)).
\bibitem{9}L. Hinvi,  A. V. Monwanou,  J. B. Chabi   Orou .,  
\emph{Linear stability  analysis of  fluid flow  between two   parallel 
porous stationary   plates  with  small suction and  injection}
 (arXiv :  $1304.7210$v$1$ [physics.flu-dyn]$12$ Apr $2013$).

\bibitem{10}J. M. McDonough.,  
\emph{Lectures in Elementary Fluid Dynamics :  Physics,  Mathematics and Applications} 
(Departments of Mechanical Engineering and Mathematics.
University of Kentucky,  Lexington,  KY $40506-0503$).
\bibitem{11}S. O. AJADI.,  
\emph {A note on Unsteady Flow of Dusty Viscous Fluid Between two parallel Plates}
(J. Appl. Math.  and Computing Vol. $18$ $(2005)$,  No. $1$ - $2$,  pp. $393 - 403$).
 \end{thebibliography}
 \end{document}